\documentclass[letterpaper]{JHEP3}

\usepackage{amsmath,amssymb}
\usepackage{bm,bbm}
\usepackage{epsfig}

\newcommand{\gsim}{\mathrel{\lower2.5pt\vbox{\lineskip=0pt\baselineskip=0pt
                   \hbox{$>$}\hbox{$\sim$}}}}
\newcommand{\lsim}{\mathrel{\lower2.5pt\vbox{\lineskip=0pt\baselineskip=0pt
                   \hbox{$<$}\hbox{$\sim$}}}}

\newcommand{\too}{\longrightarrow}

\newcommand{\abs}[1]{\left| #1 \right|}

\newcommand{\Sla}[1]{\kern0.12em{\raise.15ex\hbox{$/$}\kern-.74em #1}}

\newcommand{\beq}{\begin{eqnarray}}
\newcommand{\eeq}{\end{eqnarray}}
\newcommand{\nn}{\nonumber}
\newcommand{\eql}[1]{\label{eq:#1}}
\newcommand{\eq}[1]{(\ref{eq:#1})}

\newcommand{\del}{\partial}

\newcommand{\varep}{\varepsilon}

\title{The 't Hooft Model As A Hologram}

\author{Emanuel Katz\\
Department of Physics, Boston University,\\
590 Commonwealth Avenue, Boston, MA 02215, USA\\
E-mail: \email{amikatz@buphy.bu.edu}}

\author{Takemichi Okui\\
Department of Physics \& Astronomy, Johns Hopkins University,\\
3400 North Charles Street, Baltimore, MD 21218, USA\\
Department of Physics, University of Maryland, College Park, MD 20742, USA
\\E-mail: \email{okui@pha.jhu.edu}}

\abstract{We consider the 3d dual of $1+1$ dimensional large-$N_c$ QCD with quarks in the fundamental representation, also known as the 't Hooft model.  
't Hooft solved this model by deriving a Schr\"odinger equation for the 
wavefunction of a parton inside the meson.  In the scale-invariant limit, 
we show how this equation is related by a {\it transform} to the equation of 
motion for a scalar field in AdS$_3$.  We thus find an explicit map between 
the `parton-$x$' variable and the radial coordinate of AdS$_3$.  This direct 
map allows us to check the AdS/CFT prescription from the 2d side.  We describe 
various features of the dual in the conformal limit and to the leading order in 
conformal symmetry breaking, and make some comments on the 3d theory in the 
fully non-conformal regime.}


\preprint{arXiv:0710.3402\\UMD-PP-07-011}

\begin{document}

\section{Introduction}
\label{sec:intro}

Our limited understanding of gauge theory dynamics in the non-perturbative
regime hampers both our description of QCD phenomena, as well as our ability 
to construct viable scenarios with strong dynamics for physics beyond the
Standard Model.  Lattice theory has been helpful in addressing some of the
issues, however it does face certain challenges.  Some of these difficulties
include treatment of time evolution in a system with temperature or chemical
potential, simulation of supersymmetric theories, and dealing with chiral
symmetry in an efficient manner.  Thus, it is desirable to find novel
theoretical tools to tackle non-perturbative physics.  The AdS/CFT framework
\cite{AdS-CFT} offers a different approach for performing calculations in field
theory in the non-perturbative regime.  The local operators of the original
field theory are mapped to fields propagating in a curved higher-dimensional
background.  A general field theory contains a multitude of local operators, 
and therefore its higher-dimensional dual is expected to contain infinitely many
fields. The interactions of these higher-dimensional fields, which can be of
large spin, are expected to be quite complicated, and in general are difficult
to determine.  Considerable simplification occurs when the field theory admits 
a limit for which most of the operators acquire large anomalous dimensions.  
The anomalous dimensions are mapped via AdS/CFT to masses of the dual 
higher-dimensional fields, and thus such a limit effectively decouples most
fields.  The remaining fields are usually those dual to operators whose
dimensions are protected by various symmetries.  These are typically duals of
currents (and possibly their superpartners), and their interactions are heavily
constrained by symmetry.  Thus, most known duals are of theories where there is
a significant hierarchy between the dimensions of operators.  Unfortunately,
this is not the case for QCD, which is partly why it has thus far been difficult
to construct its dual, though duals to other field theories with `QCD-like'
dynamics have been found. In a few cases it has been possible to find soluble
higher-dimensional string duals to certain field theories (or sub-sectors
thereof)\cite{BMN}.  Such descriptions capture effectively the physics of many 
higher-dimensional fields (the resonances of the string), going beyond the
limited set constrained by symmetry.  One may hope that a theory like QCD 
admits such a string description, however thus far, none has been found.  

Hence, instead of attempting to find a dual to the full QCD theory, it
might be fruitful to consider only a limited set of operators, and find
a description for their holographic dual fields.  Such an approach faces
certain obvious challenges.  The first is that one would expect that any
operator has non-trivial correlation functions with many other operators 
(as allowed by symmetry and Lorentz invariance), and thus its dual field 
will necessarily interact with many other fields.  As mentioned, these
interactions are difficult to determine and usually are not even 
renormalizable. However, in the limit of large number of colors, $N_c$,
all interactions are suppressed, and one is left with a quadratic action of
free fields propagating on some background.  One may worry that such an 
action includes higher-derivative terms.  After all, there is no parameter
in QCD, such as the 't Hooft coupling, that would suppress them.  However,
leading $1/N_c$ calculations correspond to `on-shell' calculations in the
higher-dimensional theory, and thus only care about the dispersion relation
governing the propagation of the dual field in the curved background.  If we
know the background exactly and include {\it all} (typically an infinite number
of) fields, then in principle we can find a basis of fields where dispersion
relations become quadratic in derivatives, hence the action is local in this
sense. Thus, if we limit ourselves to asking questions that concern only the
quadratic part of the action (i.e.~focus on masses, decay constants, and 
two-point functions), this approach may be useful.  Finally, there is the
question of the curved background itself.  In the UV QCD is asymptotically 
free, and therefore the background should approach AdS. Thus, a natural place 
to start is the conformal limit of QCD, for which we know much more about the
quadratic action. Indeed, 4d Poincar\'e invariance tells us that the dispersion
relation is quadratic in 4d momenta, so derivatives with respect to 4d
coordinates enter quadratically in the action. The AdS isometry then guarantees
that derivative along the 5th coordinate also enters quadratically.  This plus
the usual consideration of internal symmetries, etc.~completely fixes the form
of the quadratic action (at least for propagating fields).  In addition, we 
need only consider the duals of primary operators as their descendents are automatically included by the AdS isometry.  As primary operators do not mix,
this is a basis of for bulk fields for which the quadratic action becomes
diagonal.

The simplicity will be lost once we take into account the effects of
conformal symmetry breaking, such as the running QCD coupling, confinement,
chiral symmetry breaking, etc.  Such effects can be parameterized in terms of
various backgrounds in the higher-dimensional space. Denoting the 5th coordinate
by $z$, these background in general depend on $z$. Then, it is no longer true that the quadratic 4d dispersion implies that $\del_z$ appears quadratically 
in the action.  For example, suppose we are interested in the quadratic action
for a scalar field $\phi$ and there is a background of another scalar field 
$\Phi$ parameterizing some conformal symmetry breaking effects.  In the 
{\it full} action, there might be a term like
$g^{M_1 N_1} g^{M_2 N_2} g^{M_3 N_3} g^{M_4 N_4}
(\del_{M_1}\Phi) (\del_{M_2}\Phi) (\del_{M_3}\Phi) (\del_{M_4}\Phi)
(\del_{N_1}\del_{N_2}\phi) (\del_{N_3}\del_{N_4}\phi)$.
Once a $z$-dependent $\Phi$ background is turned on, this yields
a {\it quadratic} term for $\phi$ with {\it four} $\del_z$'s.  Therefore,
away from the exact AdS, we do not know how many $z$ derivatives are in the
action.  Also, a term like $\Phi^2\phi^2$ will give us a $z$-dependent mass
term for $\phi$.  In addition, conformal symmetry breaking will generally induce
mixing between fields corresponding to operators with different scaling
dimensions.  But as mentioned above, these higher derivative terms are 
merely a consequence of integrating out heavier fields which mix with $\Phi$.
Once we `integrate in' all fields and include all the mixings among them,
there should be a basis for the fields for which the quadratic action is 
local.  

The above complexity means that it may be difficult to {\it derive} the dual 
of QCD but we might at least {\it learn} something about the full theory.
Restricting to a regime where QCD is almost conformal (i.e.~looking at
the correlators at large Euclidean momenta), we can match the (small) conformal
breaking effects order-by-order in $\Lambda_{\rm QCD}$.  This tells us how the backgrounds affect the quadratic Lagrangian at small $z$ (the UV of theory).
This knowledge may be sufficient for certain questions.  If for example, 
a particular bulk mode profile is localized sufficiently far from the large 
$z$ region, then the details of conformal symmetry breaking might not be very 
important in determining its properties.

The above philosophy is the motivation for the `AdS/QCD' phenomenological
approach which has been applied to fields of various spin 
\cite{original-AdS-QCD} \cite{tensor-mesons} \cite{eta-prime}.  
A good agreement of masses and decay constants with data is found.  This is an
indication that for low-lying KK-modes, both the large $N_c$ approximation works
remarkably well, and the profile of KK-modes is surprisingly well described by
assuming the background is close to AdS with a hard cutoff.  Still, it is clear
that such a description is naive as it does not capture the spectrum of the
highly excited modes, which lie on Regge trajectories.  A simple model which
captures the Regge spectrum was presented in \cite{linear-confinement}, but
its origin remains unclear.  In particular, as mentioned, once conformal symmetry is broken,  all fields dual to operators of similar quantum numbers 
are expected to mix in a complicated way.  It is therefore a mystery why this
mixing is effectively captured by the simple diagonal action of 
\cite{linear-confinement}.


In this paper we will attempt to test the AdS/QCD approach in a simpler setting
where there is some analytic control over the non-perturbative dynamics.
In 
particular, we will focus on two-dimensional QCD in the large $N_c$ limit.
The spectrum of this model was solved by 't Hooft \cite{tHooft}, 
who derived a Schr\"odinger equation for the meson wavefunction (as a function 
of the parton-$x$ variable). 
While one could ``build'' a 3d AdS/QCD model with a few fields propagating in some effective
background chosen to reproduce the meson spectrum, that is {\it not} the goal of this paper.
As mentioned above, our view is that such 3d model is an approximation
of the (quadratic) action involving an infinite number of fields mixed with each other, 
corresponding to the infinite number of operators mixed with each other on the 2d side.
Our goal is to understand such mixings and how they are mapped between 2d and 3d, taking advantage of
the exact two-point functions calculated in \cite{Gross}.

Toward this goal, we will first begin with the conformal limit of the theory where there are 
no mixings, and explicitly construct quadratic 3d actions for spin-0, -1, and -2 fields which reproduce 
the expected two-dimensional correlation functions.
This will reveal some qualitative features of the 3d actions which should be shared 
by fields with spin $\geq 3$.  We will then analyze the leading conformal symmetry breaking effects,
i.e.~the leading mixing effects, 
in particular, the chiral condensate. 
We will then return to the conformal limit and construct a ``transform'' which can directly map
the scale invariant limit of the 't Hooft equation (derived first in \cite{Rich}) to the equation 
of motion for a scalar field in AdS$_3$.  Our transform reveals an explicit
relation between the parton-$x$ variable and the radial coordinate of AdS$_3$, 
which we use to transform the meson parton wavefunction into the KK-mode
wavefunction of the dual scalar field.%
\footnote{An alternative proposal for the relation between parton-$x$ and the
radial coordinate was given in \cite{brodsky}.}
We also show how a calculation of a two-point correlator using parton
wavefunctions can be reformulated as an evaluation of an appropriate 
three-dimensional action, thereby verifying the AdS/CFT prescription.  
In other words, we find a direct map from the CFT to AdS.  

The paper is organized as follows.  In section~\ref{sec:tHooft}, we will
briefly review the 't Hooft model and summarizes the relevant results.
Section~\ref{sec:dual} which discusses the 3d dual will be divided in two parts.
In the first part, section~\ref{sec:conformal}, we will match two theories
in the conformal limit.  The second part, section~\ref{sec:conf-break},
will discuss conformal symmetry breaking to leading order in the coupling.
We then present our transform that relates the 't Hooft wavefunctions to the 
KK modes (section~\ref{sec:transform}), and show how one may derive a 3d action
from the 2d side.  Finally, we make some comments in section~\ref{sec:towards}
about the expected form of the full dual to the 't Hooft model and its relation
to the model of \cite{linear-confinement}.  We conclude in 
section~\ref{sec:conc}.

\section{The 't Hooft Model}
\label{sec:tHooft}

This section contains a short review of the 't Hooft model \cite{tHooft}
and summary of some known and new formulae relevant to our later discussions
on the 3d dual.  Section \ref{sec:basics} reviews the basic features 
of the model in the conventional language commonly used in the literature, 
while section \ref{sec:primary} and \ref{sec:tHooft-correlators} are
written in a manner best-suited for the use of AdS/CFT correspondence.  
In section \ref{sec:chiral-sym-break} we remark briefly on the fate of
chiral symmetry in the 't Hooft model.

\subsection{The Basics}
\label{sec:basics}

{\it The 't Hooft model} is an $SU(N_c)$ gauge theory in 1+1 dimensions with 
$N_{\!f}$ Dirac fermions (`quarks') in the fundamental representation
of $SU(N_c)$.  Just for simplicity, we will take $N_{\!f}=1$ in this paper.
Denoting the `quark' and the `gluon' field-strength by $\psi$ and 
$G_{\!\mu\nu}$, the Lagrangian is given by
\beq
  {\cal L}_\text{'t Hooft}
  = - \frac{N_c}{4\pi\Lambda^2} {\rm tr}\bigl[ G_{\!\mu\nu} G^{\mu\nu} \bigr]
    + i\overline{\psi} \Sla{D} \psi - m_q \overline{\psi} \psi  \>,
\eql{tHooftLagrangian}
\eeq
where $m_q$ is the quark mass, and the gluon field is normalized such 
that $D_\mu\psi=\del_\mu\psi + iA^a_\mu T^a \psi$ with 
${\rm tr}[T^a T^b] = \delta^{ab}$.  Note that in 2d the mass dimension of the
gauge coupling is one, and in \eq{tHooftLagrangian} we have chosen to write 
the coupling as $\Lambda \sqrt{\pi/N_c}$ where $\Lambda$ is a physical mass
scale analogous to $\Lambda_{\rm QCD}$ of real-life QCD.  We assume $N_c \gg 1$
and will analyze the theory in terms of $1/N_c$ expansion.  We will frequently
refer to the left-mover $\psi_+ \equiv \hat{P}_+ \psi$ and the right-mover 
$\psi_- \equiv \hat{P}_- \psi$, where $\hat{P}_\pm \equiv (1 \pm \gamma_3)/2$ 
with $\gamma_3 \equiv \gamma_0\gamma_1$. 

In this paper, we will mainly consider the $m_q \to 0$ limit, in which 
the Lagrangian \eq{tHooftLagrangian} has the following global 
$U(1)_L \otimes U(1)_R$ flavor symmetry.  Under $U(1)_L$, $\psi_+$ transforms 
as $\psi_+ \to e^{i \alpha_\ell} \psi_+$ while $\psi_-$ is neutral.  Under 
$U(1)_R$, $\psi_+$ is neutral while $\psi_-$ transforms as 
$\psi_- \to e^{i \alpha_r} \psi_-$.  Equivalently, we will sometimes talk about the vector $U(1)_V$ and axial $U(1)_A$ symmetries corresponding to 
$\alpha_\ell + \alpha_r$ and $\alpha_\ell - \alpha_r$.  Note that, 
unlike in the 4d QCD, the $SU(N_c)$ gauge interaction does {\it not} 
make $U(1)_A$ anomalous, thanks to the fact that all $SU(N_c)$ generators 
are traceless.  Therefore, in the $m_q \to 0$ limit, the Noether currents 
$L_\mu$ and $R_\mu$ for $U(1)_L$ and $U(1)_R$ are both exactly conserved even 
at quantum level. In other words, 
$\del_\mu \langle\alpha| L^\mu |\beta\rangle 
= \del_\mu \langle\alpha| R^\mu |\beta\rangle = 0$ 
for any states $|\alpha\rangle$ and $|\beta\rangle$.%
\footnote{However, they may have {\it global} anomalies, that is, products 
of currents 
(such as $\langle 0|\hat{{\rm T}}\{ L_\mu (x) \, L_\nu (y) \}|0\rangle$) 
may be only conserved up to a local term.  This is not a problem since 
these $U(1)$ symmetries are not gauged.}

Note that, in 2d, the `gluon' has no propagating degrees of freedom---it only produces instantaneous ``Coulomb'' interactions.  
Due to this and the fact that the gauge boson self-couplings vanish in light-cone gauge ($A_+=0$ or $A_-=0$),
all two-point correlation functions between color-singlet 
quark-bilinear operators can be exactly calculated at the leading order 
in $1/N_c$ expansion \cite{Gross}.  The results can be expressed solely 
in terms of the {\it 't Hooft wavefunction} $\phi_n (x)$ where $x$ is 
restricted as $0 \leq x \leq 1$ while $n=0,1,2,\cdots$ labels the mesons.
The $x$ variable is literally the $x$ in the parton model, and $|\phi_n(x)|^2$ 
is precisely the parton distribution function.  The meson mass $m_n$ is 
an eigenvalue of the {\it 't Hooft equation} (with $\phi_n(x)$ being the eigenfunction): 
\beq
  \frac{m_q^2/\Lambda^2-1}{x(1-x)} \, \phi_n (x)
    - \hat{\rm P}\!\!\int_0^1\!\! \frac{\phi_n (y)}{(y-x)^2} \, dy
  = \frac{m_n^2}{\Lambda^2} \,\phi_n (x)  \>,
\eql{tHooftEq}
\eeq
where $\hat{{\rm P}}$ denotes the principal-value prescription for the integral. 
From this equation, one can deduce that $\phi_n(x)$ can be taken to be real, orthonormal, 
and complete:
\beq
  \int_0^1\!\! dx \> \phi_n (x) \, \phi_m (x) = \delta_{nm}
    \quad, \qquad
  \sum_{n=0}^\infty \phi_n (x) \, \phi_n (y) = \delta (x-y)  \>.
\eql{orthonormal-complete}
\eeq
Also, the meson spectrum is non-degenerate, so $\phi_n(x)$ satisfies 
the following reflection property:
\beq
  \phi_n (1-x) = (-1)^n \, \phi_n (x)  \>.
\eql{phi-reflection}
\eeq

As an example which illustrates how $\phi_n(x)$ appears in the correlators, 
let us consider the scalar and pseudoscalar operators 
$S \equiv \overline{\psi} \psi$ and $P \equiv i\overline{\psi} \gamma_3 \psi$.
Then, at the leading order in $1/N_c$ expansion, the Fourier transforms of 
the $SS$ and $PP$ correlators%
\footnote{We use the notation
 \beq
   \langle {\cal O}_1 \, {\cal O}_2 \rangle (q) \equiv
    \int\! d^2x \, e^{iq\cdot x} \langle 0| {\rm \hat{T}} \{ {\cal O}_1(x) \, 
                                            {\cal O}_2 (0) \} |0 \rangle
   \nn  \>.
 \eeq} 
are given by
\beq
  \langle S \, S \rangle (q)
  &=& \frac{iN_c}{4\pi} \sum_{n=1,3,\cdots} \frac{m_q^2}{q^2 - m_n^2 + i\varep}
      \left[ \int_0^1\!\! dx \, \frac{2x-1}{x(1-x)} \phi_n (x) \right]^2
      \>, \eql{full-SS}  \\
  \langle P \, P \rangle (q)
  &=& \frac{iN_c}{4\pi} \sum_{n=0,2,\cdots} \frac{m_q^2}{q^2 - m_n^2 + i\varep}
      \left[ \int_0^1\!\! dx \, \frac{1}{x(1-x)} \phi_n (x) \right]^2
      \>. \eql{full-PP}
\eeq
(See appendix \ref{app:correlator-calc} for the derivation.)  Notice that the correlators \eq{full-SS} 
and \eq{full-PP} have poles corresponding to the meson masses, but have no cuts
associated with intermediate states of quarks---quarks are confined.
Also, we see in \eq{full-SS} and \eq{full-PP} that the $n=0,2,4,\cdots$ mesons
are pseudoscalars while the $n=1,3,5,\cdots$ mesons are scalars.

Unfortunately, no closed-form expression is known for either $\phi_n (x)$ or
$m_n$.  However, for $n \gg 1$ and $m_q \ll \Lambda$, 
it is easy to check that they may be approximated as
\beq
  \phi_n (x) \simeq \sqrt{2} \cos[n\pi x]
  \quad, \quad
  m_n^2 \simeq \pi^2\Lambda^2 n  \>.
\eql{approx-soln}
\eeq
Note that the meson spectrum exhibits a Regge-like behavior.  
This approximate form of $\phi_n (x)$ is only valid 
away from the $x=0,1$ endpoints.  Near the endpoints, $\phi_n(x)$ sharply rises from $0$ as $x^{m_q/\Lambda}$, then quickly switching to 
the above cosine behavior.%
\footnote{The reader familiar with the 't Hooft model may recognize 
that our approximate solution \eq{approx-soln} is different from the one 
commonly found in the literature where it is $\sin[(n+1)\pi x]$ instead of 
cosine.  The reason for the difference is $m_q$.  We are interested in 
the $m_q \ll \Lambda$ case (in fact the $m_q \to 0$ limit) where 
$\phi_n$ shoots up almost vertically at the endpoints because the slope 
of $x^{m_q/\Lambda}$ diverges for $m_q \to 0$.  On the other hand, 
the sine solution seen in the literature is appropriate for 
$m_q \simeq \Lambda$.}

Some exact results are known in the $m_q \to 0$ limit.  For example, we will
see in section \ref{sec:first-order-mixing} that all the mesons except $n=0$
satisfy
\beq
  \int_0^1\!\! \phi_n(x) \,dx = O(m_q/\Lambda) \too 0  \>.
\eql{int-phi_n}
\eeq
The lightest meson (i.e.~$n=0$), on the other hand, satisfies
\beq
  \phi_0 (x) \too 1
  \quad, \quad
  \frac{m_0^2}{m_q} \too \frac{2\pi}{\sqrt{3}} \Lambda  \>.
\eql{massless-properties}
\eeq
(See, for example, \cite{Neuberger} for a derivation of the last formula.)
Thus this pseudoscalar meson becomes massless as $m_q \to 0$. 
Even though this is reminiscent of the relation $m_\pi^2 \propto m_q$ in 
real-life QCD, it is actually a bit subtle to interpret the $n=0$ meson
as a Nambu-Goldstone boson from chiral symmetry breaking, because in 2d 
there is no spontaneous breaking of a continuous internal symmetry 
\cite{CMW-theorem}.  We will briefly return to this issue 
in section \ref{sec:chiral-sym-break}.

\subsection{Primary Operators in the 't Hooft Model}
\label{sec:primary}

When we construct the 3d dual of the 't Hooft model in section \ref{sec:dual},  our starting point will be the conformal limit of the model ($\Lambda \to 0$ 
and $m_q \to 0$).  In conformal field theory, primary operators play 
an important role.  Conformal invariance strongly constrains the properties 
of primary operators, and once we know all the correlation functions 
among primary operators, all other correlators can be derived from them 
by conformal symmetry.  So in this section we describe the primary operators
in the 't Hooft model.    

Since we are working at the leading order in $1/N_c$ expansion, we only consider
color-singlet quark-bilinear operators.  Furthermore, in the conformal limit,
since $m_q$ is absent and the gauge interaction can be ignored, many 
of those operators actually vanish by the equations of motion 
$\del_+\psi_- = \del_-\psi_+ = 0$.%
\footnote{The light-cone coordinates $x^\pm$ are defined as 
$x^\pm = (x^0 \pm x^1)/\sqrt{2}$.  The left-mover $\psi_+$ and 
the right-mover $\psi_-$ are defined by $\psi_\pm = \hat{P}_\pm \psi$ where 
$\hat{P}_\pm \equiv (1 \pm \gamma_3)/2$ with 
$\gamma_3 \equiv \gamma_0\gamma_1$. }
We then classify non-vanishing ones according to scaling dimensions and 
$U(1)_A$ charges.

Among $U(1)_A$-charged primary operator, the only one combination which does not vanish by the equations of motion is
\beq
  X \equiv \frac{S+iP}{\sqrt2} = \sqrt2 \psi_+^\dagger \psi_-  \>.
\eeq
All other ones can be written as a non-primary operator plus a piece
that vanishes by the equations of motion.  (See appendix \ref{app:primary-op}
for the details.)  $X$ is neutral under $U(1)_V$.  The scaling dimension of $X$
is one.       

On the other hand, there are two types of $U(1)_A$-neutral primary operators
which do not vanish by the equations of motion:
\beq
  L_{k+} &=& \sqrt{2} \sum_{j=0}^{k-1} \, ( {}_{k-1} {\rm C}_j )^2 \,
             [ (-i\del_+)^{k-1-j} \psi_+^\dagger ] \, (i\del_+)^j \psi_+
             \>, \nn\\
  R_{k-} &=& \sqrt{2} \sum_{j=0}^{k-1} \, ( {}_{k-1} {\rm C}_j )^2 \,
             [ (-i\del_-)^{k-1-j} \psi_-^\dagger ] \, (i\del_-)^j \psi_-
             \>,                  
\eql{Lk+_Rk-_def}
\eeq
where ${}_n {\rm C}_m \equiv n!/[m! \, (n-m)!]$, and the notation $L_{k+}$ is 
a shorthand for $L_{++ \cdots +}$ with $k$ $+$s.  (See appendix
\ref{app:primary-op} for derivation.)  Both the $L$-type and $R$-type are
neutral under $U(1)_V$.  The scaling dimensions of $L_{k+}$ and $R_{k-}$ are
both $k$. 

Even though $L_{k+}$ (or $R_{k-}$) by itself is an irreducible representation 
of the 2d Lorentz group, it is often convenient to regard $L_{k+}$ and $R_{k-}$
as components of the rank-$k$ tensor operators $L^{(k)}_{\mu_1 \cdots \mu_k}$ and $R^{(k)}_{\mu_1 \cdots \mu_k}$ where $L^{(k)}_{\mu_1 \cdots \mu_k}$ consists
of $\psi_+^\dagger$, $\psi_+$, and $k-1$ derivatives, while 
$R^{(k)}_{\mu_1 \cdots \mu_k}$ consists of $\psi_-^\dagger$, $\psi_-$ 
and $k-1$ derivatives.  So, {\it by definition} we have
\beq
  L^{(k)}_{++ \cdots +} \equiv L_{k+}  \quad, \qquad
  R^{(k)}_{-- \cdots -} \equiv R_{k-}  \>,
\eql{LR-def}
\eeq
{\it and}
\beq
  L^{(k)}_{-- \cdots -} \equiv 0  \quad, \qquad
  R^{(k)}_{++ \cdots +} \equiv 0  \>.
\eql{LR-zero-entry}
\eeq
All the remaining components (with mixed $+$s and $-$s) are not 
{\it identically} zero like \eq{LR-zero-entry}, but will vanish 
{\it by the conformal-limit equations of motion} $\del_+ \psi_- = \del_- \psi_+ = 0$:
\beq
  L^{(k)}_{\text{$+$$-$ mixed}} = 0  \quad, \qquad
  R^{(k)}_{\text{$+$$-$ mixed}} = 0  \quad \text{(by the e.o.m.)} 
\eql{+-mixed}
\eeq
Thus the meanings of ``0'' in \eq{LR-zero-entry} and \eq{+-mixed} are very 
different---while \eq{LR-zero-entry} is always true by definition, \eq{+-mixed}
will not hold once we go away from the conformal limit by turning on $\Lambda$
or $m_q$.  Also, even in the conformal limit, \eq{+-mixed} may be violated 
by a {\it local} term for products of operators, since quantum mechanically 
equations of motion only hold up to a local term for operator products.
   
Hereafter, we will often refer to $L^{(k)}_{\mu_1 \cdots \mu_k}$ and $R^{(k)}_{\mu_1 \cdots \mu_k}$ as `spin-$k$' currents, even though there is no angular
momentum in 1+1 dimensions.  The spin-1 and -2 currents are the familiar ones;
$L_\mu$ and $R_\mu$ are the Noether currents for $U(1)_L$ and $U(1)_R$, while
$(L_{\mu\nu} + R_{\mu\nu})/2$ is the energy-momentum tensor $T_{\mu\nu}$.
Similarly, we will sometimes refer to $X$ as `spin-0'.

\subsection{Two-Point Correlators in the 't Hooft Model}
\label{sec:tHooft-correlators}

Here, we summarize two-point correlation functions among the primary operators 
in the 't Hooft model.  We first present exact formulas at the leading order 
in the $1/N_c$ expansion (see appendix \ref{app:correlator-calc} 
for derivation), then analyze their conformal limit and 
the $O(\Lambda)$ corrections, to prepare for the construction of the 3d dual.

The $SS$ and $PP$ correlators are already presented in \eq{full-SS} and 
\eq{full-PP}.  The $LL$- and $RR$-type correlators for arbitrary $m_q$ and 
$\Lambda$ also take a rather simple form:
\beq
  \langle L_{k+} \, L_{k'+} \rangle (q)
    &=& \frac{iN_c}{\pi}
        \sum_n \frac{q_+^{k+k'}}{q^2 - m_n^2 + i\varep} M_{k,n} M_{k',n}  
        \>,\nn\\
  \langle R_{k-} \, R_{k'-} \rangle (q)
    &=& \frac{iN_c}{\pi}
        \sum_n \frac{q_-^{k+k'}}{q^2 - m_n^2 + i\varep} M_{k,n} M_{k',n}  \>,
\eql{full-O_k-O_l}
\eeq
where the {\it moments} $M_{k,n}$ are defined as
\beq
  M_{k,n} \equiv \int_0^1\!\! dx \> P_{k-1}(2x-1) \, \phi_n (x)  \>,  
\eql{moments}
\eeq
where $P_n (x)$ is the Legendre polynomial.  (Unfortunately, the correlators 
for the other components of $L^{(k)}_{\mu_1 \cdots \mu_k}$ and
$R^{(k)}_{\mu_1 \cdots \mu_k}$ with mixed $+$ and $-$ indices are difficult to
compute except in the conformal limit.   The $LR$ correlator is also difficult to calculate.)
Note that, from \eq{phi-reflection} and \eq{moments}, we see that $L_{k+}$ and
$R_{k-}$ with even $k$ create scalar mesons, while with odd $k$ they create
pseudoscalar mesons.  Then, at the leading order in $1/N_c$, this has 
a simple corollary:
\beq
  \langle S\, L^{(k)}_{\mu_1 \cdots \mu_k} \rangle (q) 
  = \langle S\, R^{(k)}_{\mu_1 \cdots \mu_k} \rangle (q) &=& 0  
    \quad \text{for $k=$ odd,}  \nn\\
  \langle P\, L^{(k)}_{\mu_1 \cdots \mu_k} \rangle (q) 
  = \langle P\, R^{(k)}_{\mu_1 \cdots \mu_k} \rangle (q) &=& 0  
    \quad \text{for $k=$ even.}
\eql{trivial-mixing}
\eeq
On the other hand, 
\beq
  \langle S\, L_{k+} \rangle (q)
  = \frac{iN_c}{2\pi} \sum_n \frac{m_q \, q_+^k}{q^2-m_n^2+i\varep} \, M_{k,n}
    \int_0^1\!\! dx\, \frac{2x-1}{x(1-x)} \phi_n(x)
    \quad \text{for $k=$ even,}
\eql{full-S-L_k}
\eeq
and
\beq
  \langle P\, L_{k+} \rangle (q)
  = \frac{N_c}{2\pi} \sum_n \frac{m_q \, q_+^k}{q^2-m_n^2+i\varep} \, M_{k,n}
    \int_0^1\!\! dx\, \frac{1}{x(1-x)} \phi_n(x)
    \quad \text{for $k=$ odd.}
\eql{full-P-L_k}
\eeq
The $SR$ correlator can be obtained from \eq{full-S-L_k} by replacing 
$q_+$ with $q_-$, while the $PR$ correlators can be obtained from 
\eq{full-P-L_k} by replacing $q_+$ with $q_-$ and put an overall $-1$.

For $k=1$, the above formulas greatly simplify in the $m_q \to 0$ limit 
(but still with arbitrary $\Lambda$).  In this limit, \eq{int-phi_n} and 
\eq{massless-properties} imply $M_{1,n} = \delta_{n,0}$, which allows us 
to evaluate \eq{full-O_k-O_l} exactly for $k=\ell=1$.  Also, recall that both
$L_-$ and $R_+$ are identically zero.  Therefore, we obtain the following 
very simple expressions:
\beq
  \langle L_\mu \, L_\nu \rangle (q) 
    &=& \frac{iN_c}{\pi} \frac{q^{L}_\mu q^{L}_\nu}{q^2 + i\varep}  \>,\nn\\  
  \langle R_\mu \, R_\nu \rangle (q) 
    &=& \frac{iN_c}{\pi} \frac{q^{R}_\mu q^{R}_\nu}{q^2 + i\varep}  
        \qquad \text{($m_q \to 0$, $\Lambda$ arbitrary),}
\eql{massless-LL-RR}
\eeq
where 
\beq
  q^L_\mu \equiv \frac{q_\mu + \epsilon_{\mu\nu} q^\nu}{2}
    \quad, \qquad
  q^R_\mu \equiv \frac{q_\mu - \epsilon_{\mu\nu} q^\nu}{2}  \>,
\eql{qLqR}
\eeq
with $\epsilon_{+-} = - \epsilon_{-+} = +1$.  (Hence,  $q_+^L = q_+$ and
$q_-^L=0$, while $q_+^R=0$ and $q_-^R=q_-$.)  How about the $LR$ correlator?
Because $L_-$ and $R_+$ are identically zero, the only (potentially) nonzero
component of the $LR$ correlator is $\langle L_+ \, R_- \rangle (q)$.  Then,
since $\langle L_+ \, R_- \rangle (q)$ is a dimensionless Lorentz scalar,    
we can write the $LR$ correlator as
\beq
  \langle L_\mu \, R_\nu \rangle (q) 
    = -\frac{iN_c}{\pi} \frac{q^{L}_\mu q^{R}_\nu}{q^2 + i\varep} \,
       f(\Lambda^2/q^2) \>,     
\eql{LR-f}
\eeq
with some function $f$.  Then, denoting the $U(1)_V$ current 
as $V_\mu = L_\mu + R_\mu$, we have
\beq
  \langle V_\mu \, V_\nu \rangle (q) 
    = \frac{iN_c}{\pi} 
      \frac{\epsilon_{\mu\alpha} q^\alpha \, \epsilon_{\nu\beta} q^\beta}   
           {q^2 + i\varep}
     +\frac{iN_c}{\pi} 
      \frac{q^{L}_\mu q^{R}_\nu + q^{R}_\mu q^{L}_\nu}{q^2 + i\varep}
      \bigl[1-f(\Lambda^2/q^2) \bigr]  \>,
\eeq
which implies
\beq
  q^\mu \langle V_\mu \, V_\nu \rangle (q) 
    = \frac{iN_c \, q_\nu}{2\pi} 
      \bigl[1-f(\Lambda^2/q^2) \bigr]  \>.
\eeq
Now, the current $V_\mu$ is classically conserved and is not anomalous either.  Then, for a product of operators such as $V_\mu (x) \, V_\nu (y)$, 
the conservation of $V_\mu$ should hold up to a local term.   Therefore, 
$f$ cannot contain a negative power of $q^2$.  On the other hand,
$f$ cannot contain a negative power of $\Lambda$ in order to have a smooth 
conformal limit.  Therefore, $f$ must be a constant, which implies that 
$\langle L_+ \, R_- \rangle (q)$ is also a constant, therefore, local.  (This
can be also easily checked by a direct calculation a la \cite{Gross}.)  While 
a choice of the constant $f$ has no effect on physics, a common choice is $f=1$
so that $\langle V_\mu V_\nu \rangle$ is identically conserved
without any contact term.  However, we instead choose $f=0$, which will be convenient for our 3d analysis in section \ref{sec:dual}.  Hence, we have
\beq
  \langle L_\mu \, R_\nu \rangle (q) = 0 
  \qquad \text{($m_q \to 0$, $\Lambda$ arbitrary).}
\eql{massless-LR}
\eeq
This has an obvious physical explanation---without $m_q$, the left and 
right movers never talk to each other, no matter what $\Lambda$ is.

\subsubsection{The Conformal Limit}

In this section we specialize the conformal limit ($\Lambda \to 0$ and 
$m_q \to 0$) of the 't Hooft model.  Let us begin with \eq{full-O_k-O_l}.
First, note that without $m_q$ or $\Lambda$ there is no dimensionful quantity 
that could make up $m_n^2$.  So we simply ignore the $m_n^2$ in the denominators
in \eq{full-O_k-O_l}, and we obtain
\beq
  \langle L_{k+} \, L_{k'+} \rangle (q)
    &=& \frac{iN_c}{\pi} \frac{\delta_{kk'}}{2k-1}
        \frac{q_+^{k+k'}}{q^2 + i\varep}  \>,\nn\\
  \langle R_{k-} \, R_{k'-} \rangle (q)
    &=& \frac{iN_c}{\pi} \frac{\delta_{kk'}}{2k-1}
        \frac{q_-^{k+k'}}{q^2 + i\varep}  
        \qquad \text{($\Lambda \to 0$, $m_q \to 0$),}
\eeq
where we have used the completeness relation of the 't Hooft wavefunctions 
\eq{orthonormal-complete} and the orthogonality of 
the Legendre polynomials.  Next, because of \eq{LR-zero-entry} and \eq{+-mixed}, 
all the remaining components of the $LL$ and $RR$ correlators are either
literally zero, or vanishing up to local terms by the equations of motion.
So let us simply set all of them to zero.  We can then summarize the $LL$ 
and $RR$ correlators in a compact form:
\beq
  \langle L^{(k)}_{\mu_1 \cdots \mu_k} \, L^{(k')}_{\nu_1 \cdots \nu_{k'}}
  \rangle (q)
  &=& \frac{iN_c}{\pi} \frac{\delta_{kk'}}{2k-1}
      \frac{q^{L}_{\mu_1} \cdots q^{L}_{\mu_k} q^{L}_{\nu_1} \cdots 
            q^{L}_{\nu_{k'}}}{q^2 + i\varep}  \>, \nn\\
  \langle R^{(k)}_{\mu_1 \cdots \mu_k} \, R^{(k')}_{\nu_1 \cdots \nu_{k'}}
  \rangle (q)
  &=& \frac{iN_c}{\pi} \frac{\delta_{kk'}}{2k-1}
      \frac{q^{R}_{\mu_1} \cdots q^{R}_{\mu_k} q^{R}_{\nu_1} \cdots 
            q^{R}_{\nu_{k'}}}{q^2 + i\varep}
      \qquad \text{($\Lambda \to 0$, $m_q \to 0$),}  
\eql{conformal-O_k-O_l}            
\eeq
where $q^L_\mu$ and $q^R_\mu$ are defined in \eq{qLqR}.  
Note that these correlators vanish for $k \neq k'$, which is consistent 
with conformal invariance which tells us that any two operators with different
scaling dimensions have a vanishing two-point correlator.
  
Next, note that the $U(1)_A$ symmetry, which is unbroken in the conformal limit,
forbids $X$ from having a nonzero two-point correlator 
with any $L^{(k)}_{\mu_1 \cdots \mu_k}$ or $R^{(k)}_{\mu_1 \cdots \mu_k}$. 
Thus we have
\beq
  \langle X \, {\cal O}_k \rangle (q) 
  = \langle X^\dagger \, {\cal O}_k \rangle (q) 
  = 0  \quad \text{for all $k$} \>,
\eql{X-O_k}
\eeq
where ${\cal O}_k = L^{(k)}_{\mu_1 \cdots \mu_k}$, 
$R^{(k)}_{\mu_1 \cdots \mu_k}$.

On the other hand, as far as the symmetries are concerned, 
$L^{(k)}_{\mu_1 \cdots \mu_k}$ and $R^{(k)}_{\mu_1 \cdots \mu_k}$ with
the same $k$ may mix with each other.  However, thanks to the fact that 
the conformal limit is a free theory, one can easily see diagrammatically that 
\beq
  \langle L^{(k)}_{\mu_1 \cdots \mu_k} \, R^{(k)}_{\nu_1 \cdots \nu_k}
  \rangle (q) = 0  \quad \text{for all $k$}  \>.
\eql{LR-conformal}
\eeq
(Here we may, if we wish, add a local term to the right-hand side, which 
of course has no effect on physics.  We choose it to be zero.)

\subsubsection{Operator Mixing at $O(\Lambda)$}
\label{sec:first-order-mixing}

In this section, we stick to the $m_q \to 0$ limit, but examine $O(\Lambda)$ 
corrections to the correlators studied in the previous section.  Fortunately, 
we are not opening Pandora's box, because dimensional analysis
and Lorentz invariance imply that the only correlators that can have nontrivial $O(\Lambda)$
pieces are $\langle S {\cal O}_k \rangle$ and $\langle P {\cal O}_k \rangle$,
where ${\cal O}_k = L^{(k)}_{\mu_1 \cdots \mu_k}$, 
$R^{(k)}_{\mu_1 \cdots \mu_k}$.  All other correlators get corrections only
starting at $O(\Lambda^2)$. 

Let us begin with the $m_q \to 0$ limit of the $PL$ correlator \eq{full-P-L_k}.
First, note that by integrating both sides of the 't Hooft equation 
\eq{tHooftEq} over $x$, we obtain 
\beq
  m_n^2 \int_0^1\!\! dx \; \phi_n (x)
  = m_q^2 \int_0^1\!\! dx \, \frac{\phi_n (x)}{x(1-x)}  \>.
\eql{m_n-m_q-swap}
\eeq
For $m_n \neq 0$, this naively seems to imply that
$\int_0^1\! \phi_n (x) \, dx = O(m_q^2) \to 0$ as $m_q \to 0$.  But this is
incorrect.  To deduce the correct $m_q$ dependence, let us look at the high
energy behavior of the $PP$ correlator \eq{full-PP}.  Since the 't Hooft model
is asymptotically free, we can use the free-quark picture to calculate the $PP$
correlator for $Q^2 \equiv -q^2 \gg \Lambda^2$, which gives
$\langle PP \rangle (q) \propto \log Q$.  On the other hand, in \eq{full-PP},
this $\log Q$ behavior must arise from summing over $n$.  Since 
$m_n^2 \propto n$ for $n \gg 1$, this can happen only if the combination
$m_q \int\! dx\, \phi_n (x)/x(1-x)$ becomes independent of $n$ for $n \gg 1$.
Returning to \eq{m_n-m_q-swap}, this means that the correct behavior must be 
$\int_0^1\! dx\, \phi_n (x) = O(m_q) \to 0$ as $m_q \to 0$.  
So, to parameterize this, let us define $\gamma_n$ via
\beq
  \frac{1}{m_q} \int_0^1\!\! dx\, \phi_n (x) = \frac{\gamma_n}{\Lambda}   
  \quad \text{as $m_q \to 0$} \>,
\eeq
for $n \neq 0$.  The $n=0$ case is an exception---recall that its behavior 
in the $m_q \to 0$ limit is given in \eq{massless-properties}.  We include 
this exception by defining $\gamma_0 = \Lambda/m_q$.
Then, in the $m_q \to 0$ limit, \eq{full-P-L_k} can be written as
\beq
  \langle P\, L_{k+} \rangle (q)
  = \frac{N_c}{2\pi} \sum_n \frac{q_+^k}{q^2-m_n^2+i\varep} 
    \frac{m_n^2}{\Lambda} \, M_{k,n} \gamma_n
    \quad \text{for $k=$ odd.}
\eql{PL_k-massless}
\eeq
Now it is manifest that the $PL$ correlator begins at ${\cal O}(\Lambda)$.  
The $PR$ correlator can be obtained from the $PL$
correlator by replacing $q_+$ with $q_-$ and multiplying an overall $-1$.  Unfortunately, there is no such simple formula for $\langle S L_{k+} \rangle$ or $\langle S R_{k-} \rangle$.

For $k=1$, the $PL$ and $PR$ correlators take especially simple forms.  
Note that \eq{massless-properties} implies $M_{1,n}= \delta_{n,0}$, and also  
recall that we have $L_- =0$ by definition.  Therefore, \eq{PL_k-massless} with $k=1$ becomes  
\beq
   \langle P \, L_\mu \rangle (q)
   = \frac{N_c}{\sqrt{3}} \frac{q^L_\mu}{q^2 + i\varep} \, \Lambda  \>,  
\eql{PL-massless}
\eeq
where $q^L_\mu$ is defined in \eq{qLqR}.  Similarly, we get
\beq
   \langle P \, R_\mu \rangle (q)
   = - \frac{N_c}{\sqrt{3}} \frac{q^R_\mu}{q^2 + i\varep} \, \Lambda  \>.   
\eql{PR-massless}
\eeq
Note that, as long as $m_q \to 0$, these two formulas are exact 
at the leading order in $1/N_c$ expansion.

\subsection{(Apparent) Chiral Symmetry Breaking}
\label{sec:chiral-sym-break}

The $O(\Lambda)$ correlators derived in the previous section seem quite
puzzling.  Notice that, by combining \eq{PL-massless} with the fact that 
$\langle S L_+ \rangle = 0$ (i.e.~the $k=1$ case in \eq{trivial-mixing}), 
we obtain $\langle X L_+ \rangle \neq 0$.  Since $X$ is charged under 
$U(1)_A$ while $L_+$ is neutral, this means that $U(1)_A$ is spontaneously broken. (There is no explicit breaking since $m_q =0$.)  Even simpler, the fact
that the scalars and the pseudoscalars are not degenerate in mass indicates 
that $U(1)_A$ is broken.  However, in two dimensions, the Coleman-Mermin-Wagner
(CMW) theorem \cite{CMW-theorem} states that there is no spontaneous breaking
of a continuous internal symmetry, in the sense that any correlation function
with a net $U(1)_A$ charge (such as $\langle X L_+ \rangle$) must vanish!
So it seems that the $1/N_c$ expansion gets the vacuum wrong or assigns 
wrong charges to the operators. 

To understand how the $1/N_c$ expansion might get the $U(1)_A$
charges wrong, imagine a 2-to-2 scattering process between, say, 
two $n=1$ mesons.  We are interested in questions about the vacuum, so
let us restrict the momenta to be much less than $O(\Lambda)$.  Then, 
the process is dominated by the exchange of the massless $n=0$ meson.  
By dimensional analysis and large-$N_c$ counting, the relevant piece of 
the effective Lagrangian schematically is
\beq
  {\cal L}_{\rm eff}
  \sim \del\phi_0 \, \del\phi_0 
       +\del\phi_1 \, \del\phi_1 + m_1^2 \, \phi_1\phi_1 
       +\frac{\Lambda^2}{\sqrt{N_c}} \phi_0 \,\phi_1 \phi_1 + \cdots  \>.
\eeq
Therefore, the amplitude ${\cal M}$ for this scattering process is given by
\beq
  {\cal M}
  \sim \left( \frac{\Lambda^2}{\sqrt{N_c}} \right)^2 
       \frac{1}{(\sqrt{m_1})^4} \frac{1}{p^2}
  \sim \frac{\Lambda^{4}}{N_c \, m_1^2 \,p^2}  \>, 
\eeq
where $(\sqrt{m_1})^4$ arises from taking into account the fact that the 
$\phi_1$ particles here are nonrelativistic, and $p \ll \Lambda$ is the
magnitude of the {\it spatial} momentum transfer in the process.  Perturbative
unitarity then requires this amplitude to be $\lsim \Lambda^2/m_1^2$, 
so this description is
actually valid only for $p \gsim p_c \equiv \Lambda/\sqrt{N_c}$.  Therefore, we
do not really know the true long-distance dynamics.  In particular, since 
$\phi_0$ gets strongly coupled to $\phi_1$ at distances of order $p_c^{-1}$,
the true state describing an $n=1$ meson is not well-approximated at all by 
the state created by the $\phi_1$ field above.  Thus, in particular, we cannot
relate the $U(1)_A$ charge of the physical $n=1$ meson to that of the 
$\phi_1$ field.  In other words, the real $n=1$ meson is
a $\phi_1$ meson accompanied by virtual $\phi_0$ mesons, 
and this `cloud' of the $\phi_0$ field effectively screens the charge of the meson.
Thus, in the $1/N_c$ expansion we do not know the charges of the mesons, hence 
we do not know if $U(1)_A$ is broken.

However, any analysis that only involves distances shorter than $O(p_c^{-1})$ 
should not care about what is going on outside the `cloud'.  In particular,
since the scale $p_c$ is much lower than $\Lambda$ for large $N_c$, we can 
trust our meson spectrum.  Also, all correlators we have calculated should be valid at energies above $O(\Lambda/\sqrt{N_c})$.  (See Ref.~\cite{witten} for
a discussion on the similar `puzzle' in the Thirring model.)  
     
For our purpose, a crucial question is whether or not the 3d dual
should exhibit this `apparent' chiral symmetry breaking.  Since loop expansion
in the 3d dual should agree order-by-order with $1/N_c$ expansion in the 2d
side, tree-level analyses in the 3d side should reproduce every aspect of the 
leading-order results in $1/N_c$ expansion in the 2d side, including things
that $1/N_c$ expansion gets `wrong'!  In fact, we will see in section 
\ref{sec:conf-break} how the 3d dual incorporates this `apparent' chiral
symmetry breaking.

\section{Aspects of the 3D Dual}
\label{sec:dual}

In this section we will construct the 3d dual of the 't Hooft model.  As we have
discussed in section \ref{sec:intro}, we will focus on two-point correlation
functions, hence our 3d Lagrangian will be just quadratic in bulk fields.  What should the 3d geometry be?  Since the 't Hooft model is asymptotically free, 
it is nearly conformal in the deep UV.  Therefore, naturally, our zeroth-order 
geometry should be AdS$_3$, corresponding to the conformal limit of the 't 
Hooft model.  Then, for $z \ll \Lambda^{-1}$, conformal symmetry breaking effects can be
parameterized as small deviations from the exact AdS$_3$, which can be analyzed 
order-by-order in $\Lambda$.  Here we should emphasize the fact that expanding
the exact correlators (the ones in section \ref{sec:tHooft-correlators}) in
powers of $\Lambda$ is {\it different} from doing perturbation theory in $g$,
despite the fact $\Lambda \propto g$.  For example, recall that 
$\langle P L_\mu \rangle \propto \Lambda$.  Clearly, we cannot get this result
from first-order perturbation in $g$---exchanging one gluon already costs
us $g^2$.  If we trace back where the $\Lambda$ comes from in section
\ref{sec:first-order-mixing}, we see that it uses information about the spectrum
(specifically the mass of the lightest meson), which cannot be understood by
perturbative expansion in $g$. 

The aim of this somewhat long section is the following.  Note that our ultimate
goal is to understand the full 3d quadratic action including all fields 
dual to the primary operators.  Those fields mix with one another, but it is 
difficult to see a priori what the mixing pattern is.  Therefore, it is useful
to study the structure of the 3d action for the fields dual to low spin 
operators.  It is also reassuring to see that our `program' works to
$O(\Lambda)$.

This section is organized as follows. 
First, in section 
\ref{sec:exact}, we discuss some exact results which are a beautiful
application of the Chern-Simons term in 3d.  Then, in section
\ref{sec:conformal}, we map the conformal limit of the 't Hooft model onto a theory in AdS$_3$, and then will analyze 
$O(\Lambda)$ conformal symmetry breaking effects in section \ref{sec:conf-break}.  
Throughout the entire section \ref{sec:dual}, we will restrict to the 
$m_q \to 0$ case, but the case with a finite quark mass clearly deserves a
separate study. 

We adopt the notation $(x^M)=(x^\mu, z)$ where $M=0,1,3$ and $\mu=0,1$, with
the AdS$_3$ metric
\beq
  ds^2 = \frac{1}{z^2} \, \eta_{MN} \, dx^M dx^N  \>,
\eql{metric}
\eeq
where $(\eta_{MN})= {\rm diag}(1,-1,-1)$.
We will raise and lower indices using $\eta_{MN}$, rather than 
$g_{MN}$, so as to make $z$ dependence always explicit.  We will work in the 
$m_q \to 0$ limit, unless otherwise stated explicitly.

\subsection{The Anomalies and the Chern-Simons Terms}
\label{sec:exact}

As we will see, there are some common features to the quadratic actions 
for the bulk fields that are dual to the $U(1)_A$-neutral primary 
operators discussed in section \ref{sec:dim-1-conformal}.  One of them is
that they all contain {\it Chern-Simons terms.}  The Chern-Simons terms are
quadratic in 3d, so they are entitled to be included in our quadratic action.
In fact, it turns out that not only they must be included for symmetry reasons,
but also they are fully responsible for generating non-trivial correlators 
between primary operators with non-zero spin, such as $L_\mu$, $R_{\mu\nu}$,
etc.  In this section, we analyze the quadratic action for the fields dual 
to $L_\mu$ and $R_\mu$, which is the simplest example that illustrates the role
played by the Chern-Simons terms.  

Recall that the correlators \eq{massless-LL-RR} and \eq{massless-LR} are
completely independent of $\Lambda$.  Since conformal symmetry breaking effects
correspond to turning on some backgrounds in the 3d bulk and deforming the
geometry away from AdS$_3$, the $\Lambda$-independence of \eq{massless-LL-RR}
and \eq{massless-LR} means that 3d calculations leading to these correlators 
must be completely insensitive to the backgrounds somehow.  So, in this section,
we would like to understand from the 3d perspective why this is so.
\footnote{There are also other exact results that are
proportional to $\Lambda$, such as \eq{PL-massless} and \eq{PR-massless}.  
Since discussing these requires some information about the conformal limit,
we will come back to them after section \ref{sec:conformal}.}

First, corresponding to the Noether currents $L_\mu$ and $R_\mu$ for the 
$U(1)_L \otimes U(1)_R$ global symmetry, we introduce 3d gauge fields 
${\cal L}_M$ and ${\cal R}_M$ for the $U(1)_L \otimes U(1)_R$ gauge
symmetry.  The values of the bulk gauge fields at the $z=0$ boundary, 
$\ell_\mu(x) \equiv {\cal L}_\mu (x,0)$ and 
$r_\mu(x) \equiv {\cal R}_\mu (x,0)$, are identified as the sources for the 2d 
operators $L_\mu$ and $R_\mu$.  We then perform 3d path integral for fixed
$\ell(x)$ and $r(x)$ to obtain an effective action which is a functional of 
$\ell(x)$ and $r(x)$.  This effective action is then interpreted as the 2d
generating functional $W[\ell, r]$, from which we can obtain any correlation
functions involving $L_\mu$ and $R_\mu$.  Following our general philosophy,
we only consider two-point correlators, and in this section we restrict our
attention to two-point correlators between $L_\mu$ and $R_\mu$ only, namely,
\eq{massless-LL-RR} and \eq{massless-LR}.  We first consider the $LL$ and $RR$
correlators \eq{massless-LL-RR}, i.e.~the effective action $W_{LL}[\ell]$ and
$W_{RR}[r]$, where $W_{LL}[\ell]$ is a quadratic functional of only $\ell(x)$,
and likewise for $W_{RR}[r]$.

The key is to look at the anomalies of the $LL$ and $RR$ correlators.
Even though $L_\mu$ and $R_\mu$ are both conserved classically, taking the
divergence of \eq{massless-LL-RR} gives
\beq
  q^\mu \langle L_\mu  L_\nu \rangle (q) 
    = \frac{iN_c}{4\pi} (q_\nu + \tilde{q}_\nu)
  \quad, \qquad 
  q^\mu \langle R_\mu R_\nu \rangle (q) 
    = \frac{iN_c}{4\pi} (q_\nu - \tilde{q}_\nu)  \>,  
\eql{LL-RR-anomalies}
\eeq
where $\tilde{q}_\nu \equiv \epsilon_{\nu\rho} \, q^\rho$.  It is important to
note that no terms in \eq{LL-RR-anomalies} can be adjusted by adding local terms 
to the right-hand sides of \eq{massless-LL-RR}.  For example, naively, it 
might seem that we could add to $\langle L_\mu L_\nu \rangle$ a local term
$-i\eta_{\mu\nu} N_c/(4\pi)$ to cancel the $q_\nu$ term appearing in 
$q^\mu\langle L_\mu L_\nu\rangle$.  However, with such a local term, 
$\langle L_\mu L_\nu \rangle$ would not vanish when $\mu$ or $\nu$ is $-$,
which contradicts with the fact that there is no $L_-$.  On the other hand, 
a local term that would shift the coefficient of the $\tilde{q}_\nu$ term 
would have to be proportional to $\epsilon_{\mu\nu}$, which is impossible,
however, because $\langle L_\mu L_\nu \rangle(q)$ must be symmetric under
$\mu \leftrightarrow \nu$ and $q \to -q$.  Therefore, since the nonzero
divergences \eq{LL-RR-anomalies} cannot be cancelled by adding local terms to
$\langle L_\mu L_\nu \rangle$ or $\langle R_\mu R_\nu \rangle$, 
\eq{LL-RR-anomalies} represent anomalies of these correlators.
 
This then implies that, under 
$\ell_\mu(x) \to \ell_\mu(x) + \del_\mu \xi_\ell(x)$, $W_{LL}[\ell]$ changes as
\beq
  W_{LL}[\ell] 
    &\too& W_{LL}[\ell] + \int\! \frac{d^2q}{(2\pi)^2} \, \ell^\mu (-q) \,
                     \langle L_\mu L_\nu \rangle(q) \> q^\nu \xi_\ell (q)
                       \nn\\
       &=& W_{LL}[\ell] + \frac{iN_c}{4\pi} \int\! \frac{d^2q}{(2\pi)^2} \, 
                     \ell^\mu (-q) \, ( q_\mu + \tilde{q}_\mu ) \,
                     \xi_\ell (q)  \>.
\eql{W-change}
\eeq
On the other hand, in the 3d side, we have the $U(1)_L$ gauge transformation
\beq
  {\cal L}_M (x,z) \to {\cal L}_M (x,z) + \del_M \xi_\ell (x,z)  \>,
\eql{U(1)_L-trans}
\eeq
where
$\xi_\ell(x,0) = \xi_\ell(x)$.  The variation \eq{W-change} then clearly shows
that the 3d Lagrangian for ${\cal L}_M$ must contain a term other than the
kinetic term ${\cal F}^{(L)}_{MN} \, {\cal F}^{(L)MN}$.  
The non-invariance cannot be due to a mass 
term in the bulk, however; Such a mass term can only arise from the Higgs 
mechanism in the bulk, which would correspond to the (apparent) chiral 
symmetry breaking discussed in section \ref{sec:chiral-sym-break}, but the 
correlators \eq{massless-LL-RR} contain no $\Lambda$ and thus do not see the 
(apparent) chiral symmetry breaking.  Therefore, the gauge symmetry must be 
intact in the bulk, and it may be violated only by the presence of the 
boundary.  Then, it is easy to see that the $q_\mu$ term of \eq{W-change} must 
be reproduced by a boundary mass term 
$-\frac{N_c}{8\pi} \,{\cal L}_\mu {\cal L}^\mu$ at $z=0$.
Put another way, recall that the $q_\mu$ term would be absent if we added 
a local term that violates the identity $L_-=0$.  Therefore, the above boundary
mass term is telling AdS$_3$ that there is no such thing as $L_-$.

What about the $\tilde{q}_\mu$ term?  Since it has an $\epsilon$ tensor in it,
the only possible quadratic term in the bulk is the Chern-Simons term 
$\frac{N_c}{4\pi} \, \epsilon^{LMN} {\cal L}_L \del_M {\cal L}_N$ 
($\epsilon^{013}=+1$).  Under the $U(1)_L$ gauge transformation 
\eq{U(1)_L-trans}, this is invariant {\it up to a total derivative} which
precisely yields the boundary term we want to match the $\tilde{q}_\mu$ term in
\eq{W-change}!  Repeating the same analysis for $R_\mu$ leads to the same
coefficient for the ${\cal R}_M$ boundary mass term, while the opposite-sign
coefficient for the ${\cal R}_M$ Chern-Simons term, due to the opposite signs in
\eq{LL-RR-anomalies}.  Thus, we have exactly determined the part of the 3d
action responsible for the anomalies of the correlators \eq{massless-LL-RR} 
and \eq{massless-LR}: 
\beq
  {\cal S}_{L,R} 
   = {\cal S}_{L,R}^{\rm bulk}
    &+& \frac{N_c}{4\pi} \int\! d^2x\, dz \,     
           \epsilon^{LMN} 
           \bigl( {\cal L}_L \, \del_M {\cal L}_N
                 - {\cal R}_L \, \del_M {\cal R}_N \bigr)  \nn\\
    &-& \frac{N_c}{8\pi} \int\! d^2x \,  
           \bigl[ {\cal L}_\mu {\cal L}^\mu 
                 + {\cal R}_\mu {\cal R}^\mu \bigr]_{z=0}  \>,
\eql{S_LR}
\eeq
where ${\cal S}_{L,R}^{\rm bulk}$ refers to gauge-invariant bulk terms (such as 
the kinetic terms for ${\cal L}_M$ and ${\cal R}_M$), which do not
contribute to the divergence of the $LL$ and $RR$ correlators.

There are a few key things to notice here.  First, the Chern-Simons and the
boundary mass terms are both completely insensitive to the bulk geometry or
any background turned on in the bulk.  This is obviously true for the boundary
terms.  The Chern-Simons term is insensitive to the bulk geometry, simply 
because the metric never appears there.  Furthermore, its gauge invariance (up
to a total derivative) forbids a $z$-dependent background to multiply
${\cal L}_L \del_M {\cal L}_N$.  Therefore, nothing can feel a source of
conformal symmetry breaking, hence the divergence of the $LL$ and $RR$
correlators \eq{LL-RR-anomalies} must be exactly correct even in the presence
of $\Lambda$.

This in turn implies the following.  Note that the correlators 
\eq{massless-LL-RR} are {\it unique} once the divergences 
\eq{LL-RR-anomalies} are given.  Therefore, even without knowing anything 
about ${\cal S}_{L,R}^{\rm bulk}$, we know that the 3d side will give the
correct $LL$ and $RR$ correlators regardless of the bulk geometry or other
backgrounds turned on in the bulk!  From the 3d perspective, this is 
nontrivial because once we turn on $\Lambda$ all bulk fields mix with one
another.  We will explicitly see in section \ref{sec:dim-1-conformal} how the
3d side `knows' that the conformal result is actually exact.
 
There is also a nice interpretation of the different choices of
$f$ in \eq{LR-f} on the 3d side.  Note that the boundary terms above correspond to our
particular choice of $f$, namely, $f=0$.  If we choose $f=1$ instead so as to
have $q^\mu \langle V_\mu \, V_\nu \rangle = 0$ without any contact term,
repeating the above exercise tells us that there should be an additional mass
term $\frac{N_c}{4\pi} {\cal L}_\mu {\cal R}^\mu$ at the $z=0$ boundary in order
to match the nonzero divergence of the $LR$ correlator \eq{LR-f}.
Note that this new mass term plus the existing ones amount to 
a mass term ${\cal A}_\mu {\cal A}^\mu$ for 
${\cal A}_M \equiv {\cal L}_M - {\cal R}_M$.
Similarly, a new Chern-Simons term must be added as well, which
together with the old ones becomes a single term $\epsilon^{LMN} {\cal A}_L \del_M {\cal V}_N$.  
This is the 3d manifestation
of the well-known fact that any $U(1)_V$-preserving counterterm necessarily
violates $U(1)_A$.

\subsection{The Conformal Limit}
\label{sec:conformal}

As we have seen, the 3d action for the fields dual to
the $U(1)_A$-neutral primary operators ${\cal L}_{\mu_1 \cdots \mu_k}$ and 
${\cal R}_{\mu_1 \cdots \mu_k}$ has the feature that in the 
conformal limit it is essentially governed by the Chern-Simons term.  In 
section \ref{sec:dim-1-conformal} and \ref{sec:dim-2}, we will study the spin-1 and
-2 cases in detail and verify this feature.  We then remark on the general
structures for higher spin cases in section \ref{sec:higher-spin}, and analyze
the spin-0 case in section \ref{sec:scalar-conformal}, which in the conformal 
limit is just a standard AdS/CFT calculation.

\subsubsection{The Spin-1 Sector}  
\label{sec:dim-1-conformal}

This sector consists of operators $L_\mu$ and $R_\mu$.  In the conformal 
limit, the quadratic part of the Lagrangian is given by \eq{S_LR}
with ${\cal S}_{L,R}^{\rm bulk}$ being just the kinetic terms for ${\cal L}_M$ 
and ${\cal R}_M$ in the AdS$_3$ background:
\beq
  {\cal S}_{L,R}^{\rm bulk}
  = \int\! d^2x\, dz\, 
      \biggl[ -\frac{z}{4g_3^2} {\cal F}^{\rm (L)}_{MN} {\cal F}^{{\rm (L)}MN} 
              -\frac{z}{4g_3^2} {\cal F}^{\rm (R)}_{MN} {\cal F}^{{\rm (R)}MN}
      \biggr]  \>,
\eql{S_LR-bulk}
\eeq
where $g_3$ is the gauge coupling which is chosen to be the same for 
${\cal L}_M$ and ${\cal R}_M$ because the 't Hooft model respects parity.
First, since ${\cal L}_M$ and ${\cal R}_M$ do not couple in the Lagrangian,
the correlator \eq{massless-LR} is trivially reproduced.  Next, as we have
pointed out, the 3d side should give us the exact $LL$ and $RR$
correlators to all orders in $\Lambda$.  Since the correlators
\eq{massless-LL-RR} have no $\Lambda$, this actually means that the 3d result
should only depend on the fact that the background is asymptotically AdS$_3$,
i.e., the bulk Lagrangian can be anything as long as it asymptotically takes 
the form \eq{S_LR-bulk} as $z \to 0$.  Let us see how this comes out.

Since the Lagrangian for ${\cal L}_M$ and that for ${\cal R}_M$ are the
same except for the sign of the Chern-Simons term, let us look at 
${\cal L}_M$.
We choose a gauge where ${\cal L}_3=0$.  Furthermore,
it is convenient to decompose ${\cal L}_\mu (q,z)$ (where $q$ is the 2d 
momentum) into its longitudinal and transverse components:
\beq
  {\cal L}_\mu = \frac{iq_\mu}{q^2} {\cal L}_\parallel 
                 + \frac{i\epsilon_{\mu\nu} q^\nu}{q^2} {\cal L}_\perp  \>,
\eeq
where ${\cal L}_\parallel$ is the longitudinal component, 
i.e.~$\del^\mu {\cal L}_\mu = {\cal L}_\parallel$, while ${\cal L}_\perp$ the 
transverse. The constraint equation arising from varying the Lagrangian with
respect to ${\cal L}_3$ and setting ${\cal L}_3 = 0$ is
\beq
  \frac{1}{g_3^2} z {\cal L}'_\parallel + \frac{N_c}{2\pi} {\cal L}_\perp
  = 0  \>, 
\eql{L-constraint-conformal}
\eeq
where the prime denotes a $z$ derivative, and the coefficients should make 
clear the origin of each term.
The equation of motion in the bulk for an Euclidean momentum $Q^2 \equiv -q^2>0$
is
\beq
  \frac{1}{g_3^2} \bigl[ z(z {\cal L}_\perp')' - Q^2 z^2 {\cal L}_\perp \bigr]
   + \frac{N_c}{2\pi} z {\cal L}'_\parallel = 0  \>. 
\eql{L-EOM-conformal}
\eeq
The solution to these equations which vanishes as $z \to \infty$ are
\beq
  {\cal L}_\perp(q,z) 
    = \frac{K_\nu (Qz)}{K_\nu (Q\epsilon)} \, 
      {\cal L}_\perp(q,\epsilon)  \>,
\eql{L_perp-sol}
\eeq
where $K_\nu (x)$ is the modified Bessel function of the second kind with 
$\nu \equiv g_3^2 N_c/(2\pi)$.  Note that we have introduced a 
short-distance cutoff by moving the boundary to $z=\epsilon >0$. 
Repeating this exercise for ${\cal R}_M$ is a trivial task.  

Now, upon 
plugging the solutions into the action, there is an important intermediate 
step which provides a crucial insight.  Regarding $z$ as ``time'', we find
that the action as a functional of the ``initial conditions'' at
$z=\epsilon$ takes the following form for any ${\cal L}$ and ${\cal R}$ 
that vanish at $z=\infty$:   
\beq
  {\cal S}_{L,R}
   &=& -\int\! \frac{d^2q}{(2\pi)^2} \biggl[ 
       \frac{\epsilon}{2 g_3^2 Q^2}
       \bigl\{ {\cal L}_\perp (-q) \, {\cal L}'_\perp (q) 
              + {\cal R}_\perp (-q) \, {\cal R}'_\perp (q) \bigr\}  \nn\\
       && \qquad \qquad
       + \frac{N_c}{4\pi Q^2} 
         \bigl\{ {\cal L}_\parallel (-q) \, {\cal L}_\perp (q)
                - {\cal R}_\parallel (-q) \, {\cal R}_\perp (q)
         \bigr\}  \nn\\
       && \qquad \qquad
       + \frac{N_c}{8\pi} \bigl\{ {\cal L}_\mu (-q) \, {\cal L}^\mu (q) 
                                + {\cal R}_\mu (-q) \, {\cal R}^\mu (q) \bigr\} 
       \bigg]_{z=\epsilon}  \>, 
\eql{S_LR-surface}
\eeq
where everything is evaluated at $z=\epsilon$.
Note that, since $K_\nu (Qz) \propto z^{-\nu}$ for small $z$, we have
$\epsilon {\cal L}'_\perp (q, \epsilon) = -\nu {\cal L}_\perp (q, \epsilon) 
+ O(\epsilon)$.  
Now we can take the $\epsilon \to 0$ limit, and in terms of the original
${\cal L}_\mu$ and ${\cal R}_\mu$ variables, we get
\beq
  {\cal S}_{L,R}
   = -\frac{N_c}{2\pi} \int\! \frac{d^2q}{(2\pi)^2} \,
      \biggl[ {\cal L}^\mu (-q) \, \frac{q^L_\mu q^L_\nu}{q^2 + i\varep} \, 
                {\cal L}^\nu (q)
             + {\cal R}^\mu (-q) \, \frac{q^R_\mu q^R_\nu}{q^2 + i\varep} \, 
                 {\cal R}^\nu (q) \biggr]_{z=0}  \>,
\eql{3d-LL-RR}
\eeq
where we have analytically-continued back to the Minkowski momentum.  This 
effective action exactly gives \eq{massless-LL-RR} regardless of the value of
$g_3$, as we have expected.

In the above derivation, one should observe that the action
was dominated by the {\it leading} small-$z$ behaviors of ${\cal L}_\perp$ and ${\cal R}_\perp$.  (The only property of $K_\nu (Qz)$ that was actually used is that it behaves as $z^{-\nu}$ for small $z$.)  This means that the effective action 
\eq{3d-LL-RR} is actually completely
insensitive to the breaking of conformal invariance, because the {\it leading}
small-$z$ behavior is fixed by the requirement that the theory be  asymptotically AdS$_3$ for small $z$, reflecting the asymptotic freedom of the 't Hooft model.  
Therefore, the 3d dual also knows that the correlators \eq{massless-LL-RR} are exact!

\subsubsection{The Spin-2 Sector and the Gravitational Chern-Simons Term}
\label{sec:dim-2}

In this sector,  we have the operators $L_{\mu\nu}$ and $R_{\mu\nu}$,
as discussed in section \ref{sec:primary}.  Even though the spin-2 case 
has the same feature as spin-1 that the Chern-Simons term completely governs
the conformal limit, there is an important difference; while the conformal
result is actually exact in the spin-1 case, it will receive $\Lambda$ 
dependent corrections for spin-2 and higher.  Therefore, the spin-2 case 
serves as a `prototype' for all higher spin cases, exhibiting all the common
qualitative features and complexities. 

Setting $k=2$ in
\eq{conformal-O_k-O_l}, the correlators in the conformal limit are
\beq
  \langle L_{\mu\nu} \, L_{\rho\sigma} \rangle (q)
  &=& \frac{iN_c}{3\pi} 
      \frac{q^L_\mu q^L_\nu q^L_\rho q^L_\sigma}{q^2 + i\varep}  \>, \nn\\
  \langle R_{\mu\nu} \, R_{\rho\sigma} \rangle (q)
  &=& \frac{iN_c}{3\pi} 
      \frac{q^R_\mu q^R_\nu q^R_\rho q^R_\sigma}{q^2 + i\varep}  \>.
\eql{spin-2-LL-RR}
\eeq
We also have $\langle L_{\mu\nu} \, R_{\rho\sigma} \rangle (q) = 0$
from \eq{LR-conformal}.  In the full interacting theory, the linear combination
$(L_{\mu\nu} + R_{\mu\nu})/2$ is {\it the} energy-momentum tensor which is
conserved.  However, in the conformal limit, $L_{\mu\nu}$ and $R_{\mu\nu}$ 
are separately conserved.  Correspondingly, in the 3d side, there must be two
`gravitons', ${\cal L}_{MN}$ and 
${\cal R}_{MN}$, where {\it the} graviton is the combination
${\cal L}_{MN} + {\cal R}_{MN}$.  

Below, we begin with some formalisms concerning spin-2 fields, in
particular, the gravitational Chern-Simons term \cite{Deser}.  Then, following 
a similar
path as the spin-1 case, we first match anomalies and fix the coefficients of
the Chern-Simons terms, then we will derive the correlators, and find that
the correlators are already fixed by the Chern-Simons, that is, the 3d
predictions of $\langle L_{\mu\nu} \, L_{\rho\sigma} \rangle$ and 
$\langle R_{\mu\nu} \, R_{\rho\sigma} \rangle$ turn out to be completely
independent of the value of $M_*$ (i.e.~the 3d Planck scale).  These are
completely parallel to the spin-1 case.  But we will also see where differences
come in once we turn on $\Lambda$.

First, some generalities.%
\footnote{In this section, we distinguish two types of indices.
When an index is $L,M,N,\cdots$ (or $\mu,\nu,\rho,\cdots$ when referring to
only 2d coordinates), it is raised and lowered using $\eta_{MN}$, 
which is the convention used in all other sections in the paper.  
On the other hand, when an index is $A,B,C,\cdots$ (or 
$\alpha,\beta,\gamma,\cdots$ when referring only to the 2d coordinates), it is
raised and lowered using the honest AdS$_3$ metric $\hat{g}_{AB}$.
The spacetime covariant derivative $\nabla_{\!\!A}$ is covariant with respect
to the AdS$_3$ background $\hat{g}_{AB}$ (i.e.~{\it not} including the 
fluctuations $h_{AB}$), unless otherwise noted.}
We write the full metric $g_{AB}$ as
\beq
  g_{AB} = \hat{g}_{AB} + h_{AB}  \>,
\eeq
where $\hat{g}_{AB}$ is the background AdS$_3$ metric, and $h_{AB}$ is the
fluctuation around the background.  (Later when we apply the formalism to
our problem, $h_{AB}$ will be ${\cal L}_{AB}$ or ${\cal R}_{AB}$.)
Then, general covariance is equivalent to gauge invariance under the following
transformation of $h_{AB}$: 
\beq
  h_{AB} \too 
    h_{AB} + \nabla_{\!\!A} \xi_B + \nabla_{\!B} \xi_A  \>,
\eql{grav-gauge-trans-full}
\eeq
where $\xi^A$ is an infinitesimal transformation parameter, and terms of 
$O(\xi h)$ or higher are dropped.
In our coordinates \eq{metric}, this becomes 
\beq
  \delta h_{33} &=& \frac{2}{z} (z \xi_3)'  \>,\\ 
  \delta h_{3\alpha} 
                &=& \del_\alpha \xi_3 + \frac{1}{z^2} (z^2 \xi_\alpha)'
                    \>,\\
  \delta h_{\alpha\beta} 
                &=& \del_\alpha \xi_\beta + \del_\beta \xi_\alpha 
                    + \frac{2}{z} \eta_{\alpha\beta} \xi_3  \>,  
\eeq
where the primes denote a $z$ derivative. It allows us to choose a gauge where
\beq
  h_{33} = h_{3\alpha} = 0  \>.
\eql{grav-gauge}
\eeq
This does not completely fix the gauge, however, and the (useful part of) 
residual gauge transformations which preserve the $h_{3A}=0$ gauge can be
parameterized as
\beq
  \xi_\alpha (x,z) = \frac{1}{z^2} \tilde{\xi}_\alpha (x)  \quad,\qquad
  \xi_3 = 0  \>,
\eql{xi-tilde}
\eeq
where $\tilde{\xi}_\alpha (x)$ is independent of $z$.
Then, in terms of $\tilde{h}_{\alpha\beta}$ defined via
\beq
  h_{\alpha\beta} \equiv \frac{1}{z^2} \tilde{h}_{\alpha\beta}  \>,
\eeq
the residual gauge transformation reads
\beq
  \tilde{h}_{\alpha\beta} (x,z) 
    \too \tilde{h}_{\alpha\beta} (x,z) + \del_\alpha \tilde{\xi}_\beta (x)
                                       + \del_\beta \tilde{\xi}_\alpha (x)  \>.
\eql{grav-gauge-trans-residual}
\eeq
Note that the shift of $\tilde{h}_{\alpha\beta}$ is independent of $z$.
In other words, the zero mode (i.e.~the $z$-independent mode) of 
$\tilde{h}_{\alpha\beta}$ transforms exactly like the `graviton' in flat 2d
space.%
\footnote{The rest of the residual gauge transformation takes the form
$\xi_\alpha = - \frac12 \del_\alpha \zeta(x)$, 
$\xi_3 = \frac{1}{z} \zeta(x)$, and
$\tilde{h}_{\alpha\beta} \to \tilde{h}_{\alpha\beta} 
- z^2 \del_\alpha \del_\beta \zeta(x) + 2 \eta_{\alpha\beta} \zeta(x)$.
At the $z=0$ boundary with $h_{MN} = {\cal L}_{MN}$, this gauge transformation gives $\langle L^\mu_{~\mu} L_{\rho\sigma} \rangle$, but this is unphysical 
because it can be set to zero by adding local terms to 
$\langle L_{\mu\nu} L_{\rho\sigma} \rangle$.}

Now, at the quadratic order in $h_{AB}$, the usual Einstein-Hilbert term 
plus the cosmological constant is equal (neglecting total derivatives) to 
\beq
  {\cal L}_{\rm EH}
  &=& M_* \biggl[ \frac14 (\nabla_{\!\!A} h_{BC}) \, \nabla^A h^{BC}  
                 -\frac12 (\nabla_{\!\!A} h_{BC}) \, \nabla^B h^{AC}  
                 +\frac12 (\nabla_{\!\!A} h) \, \nabla_{\!B} h^{AB} 
                 -\frac14 (\nabla_{\!\!A} h) \, \nabla^A h  \nn\\
                    && \hspace{2EM}
                 +\frac{h^2}{2} - h^{AB} h_{AB}
          \biggr]  \>,
\eql{L_EH-full}
\eeq
where $h \equiv h^A_A$ and $M_*$ is the 3d Planck scale.  The last two terms
look like `mass' terms, but they are actually required by gauge invariance.
In fact, under the full gauge transformation \eq{grav-gauge-trans-full}, 
${\cal L}_{\rm EH}$ transforms as
\beq
  {\cal L}_{\rm EH} 
    &\too& {\cal L}_{\rm EH}  
           + M_* \nabla_{\!\!A} \biggl[ h \xi^A - h^{AB} \xi_B
              + (\nabla_{\!B} \xi^A) \nabla_{\!C} h^{BC}
              - (\nabla_{\!B} \xi^C) \nabla_{\!C} h^{AB}  \nn\\
                  && \hspace{6EM}
              + \frac12 (\nabla_{\!B} h) 
                  \bigl( \nabla^A \xi^B - \nabla^B \xi^A \bigr)
             \biggr]  \>, 
\eql{L_EH-gauge-trans}
\eeq
so it is gauge invariant up to a total derivative.
In our coordinates \eq{metric} and gauge \eq{grav-gauge}, the action from the 
Lagrangian \eq{L_EH-full} becomes
\beq
   {\cal S}_{\rm EH}
  &=& M_* \!\int\! d^2x\, \frac{dz}{z} \biggl[ 
        \frac14 (\del_M \tilde{h}_{\nu\rho}) \, \del^M \tilde{h}^{\nu\rho}  
       -\frac12 (\del_\mu \tilde{h}_{\nu\rho}) \, \del^\nu \tilde{h}^{\mu\rho}  
       +\frac12 (\del_\mu \tilde{h}) \, \del_\nu \tilde{h}^{\mu\nu} 
       -\frac14 (\del_M \tilde{h}) \, \del^M \tilde{h} 
      \biggr]  \nn\\
      &&  
      - M_* \!\int\! d^2x \, \frac{1}{\epsilon^2}
        \biggl[ \frac12 \tilde{h}_{\mu\nu} \tilde{h}^{\mu\nu} 
               -\frac14 \tilde{h}^2 \biggr]_{z=\epsilon}  \>,   
\eql{S_EH}
\eeq
where $\tilde{h} \equiv \tilde{h}^\mu_\mu$.
Note that there are no longer `mass' terms in the bulk, while boundary mass
terms have appeared at $z=\epsilon$.  Although they diverge as $\epsilon \to 0$,
they are merely local, thus we simply throw them away.  Then, 
${\cal S}_{\rm EH}$ will be completely invariant under the residual gauge
transformation \eq{grav-gauge-trans-residual}.  (Hereafter, when we refer to 
\eq{S_EH}, the last two terms at $z=\epsilon$ will not be included.)
  
On the other hand, the gravitational Chern-Simons term can be constructed 
by a direct analogy with the Chern-Simons term for a non-Abelian gauge field
\cite{So-Young}.  We define ${\bf \Gamma}_{\!A}$ to be a matrix whose 
${}^B_{~C}$-component is equal to the Christoffel coefficient $\Gamma^B_{~AC}$,
that is, $({\bf \Gamma}_{\!A})^B_{~C} \equiv \Gamma^B_{~AC}$.  Similarly, we
define $\hat{R}_{AB}$ to be a matrix whose components are given by the Riemann
tensor $R^C_{~DAB}$, that is, $({\bf R}_{AB})^C_{~D} \equiv R^C_{~D AB}$.  For
example, in this notation, we have
\beq
  {\bf R}_{AB} = \bigl[ \del_A + {\bf \Gamma}_{\!A} ,\, 
                        \del_A + {\bf \Gamma}_{B} \bigr]  \>,
\eql{nabla-commu}
\eeq
so ${\bf \Gamma}_{\!A}$ and ${\bf R}_{AB}$ are exactly analogous to
a non-Abelian gauge field $A_A$ and its field-strength $F_{AB}$.  Then, from 
the form of the Chern-Simons term for the non-Abelian gauge field,
$\epsilon^{ABC} \, {\rm Tr} \bigl[ \frac12 A_A F_{BC} 
- \frac13 A_A A_B A_C \bigr]$, we can immediately write down 
the gravitational Chern-Simons term $\Omega_{\rm CS}$:
\beq
  \Omega_{\rm CS} 
    = \epsilon^{ABC} \, 
      {\rm Tr} \biggl[ \frac12 {\bf \Gamma}_{\!A} {\bf R}_{BC} 
                     - \frac13 {\bf \Gamma}_{\!A} {\bf \Gamma}_{B} 
                       {\bf \Gamma}_{C} \biggr]  \>. 
\eql{Omega_CS}
\eeq
Under the gauge transformation \eq{grav-gauge-trans-full}, 
the gravitational Chern-Simons form \eq{Omega_CS} transforms as
\beq
  \Omega_{\rm CS} 
    \too \Omega_{\rm CS} + \del_{\!A} (\xi^A \Omega_{\rm CS})
         + \epsilon^{ABC} (\del_{\!A} \del_D \xi^E) \, 
           \del_B \Gamma^D_{CE}  \>. 
\eql{Omega_CS-gauge-trans}
\eeq
Then, the action for $h_{AB}$ is ${\cal S}_{\rm EH}+{\cal S}_{\rm CS}$ where
\beq
  {\cal S}_{\rm CS}
  \equiv  c \!\int\! d^2x\, dz\, \Omega_{\rm CS}  \>,
\eeq
with a constant $c$ to be determined below.  
In our coordinates \eq{metric} and gauge \eq{grav-gauge}, this becomes
\beq
  {\cal S}_{\rm CS} 
  = c \int\! d^2x\,dz \, \epsilon^{\mu\nu} \left[ 
        \frac{1}{2} (\partial_{\rho} \tilde{h}_{\mu \sigma}) 
          ( \partial^{\rho} \tilde{h}_{\nu}^{\prime\sigma}
           -\partial^{\sigma} \tilde{h}_{\nu}^{\prime\rho}) 
           -\frac{1}{2} \tilde{h}'_{\mu\rho} \tilde{h}_{\nu}^{\prime\prime\rho})
      \right]  \>, 
\eql{S_CS}
\eeq
while the gauge transformation \eq{Omega_CS-gauge-trans} reduces to
the following boundary term at $z=0$:
\beq
  \delta {\cal S}_{\rm CS} 
  = -\frac{ic}{2} \int\! \frac{d^2q}{(2\pi)^2} \, \tilde{\xi}^\nu (-q) \, 
    \tilde{q}_\nu \tilde{q}_\rho \tilde{q}_\sigma \, \bar{h}^{\rho\sigma}(q)
      \>,   
\eql{CS-boundary}
\eeq
where $\tilde{q}_\mu \equiv \epsilon_{\mu\nu} q^\nu$, and
$\bar{h}^{\mu\nu} (q) \equiv \tilde{h}^{\mu\nu} (q,z)|_{z=0}$.  (Note that
$\tilde{\xi}_\mu$ is defined in \eq{xi-tilde}; it is not
$\epsilon_{\mu\nu} \xi^\nu$.) 

We now apply the formalism to the construction
of the 3d dual of the $L_{\mu\nu}$-$R_{\mu\nu}$ sector.  
Since $\langle L_{\mu\nu} R_{\rho\sigma} \rangle =0$ in the conformal limit,
and the difference between the $LL$ and $RR$ sectors are trivial sign
differences, we consider the $LL$ correlator below, and point out whenever
there is a sign difference for the $RR$ case.  The following calculations
can be divided in two parts; the first part is analogous to the analysis in
section \ref{sec:exact} where we match anomalies and fix the normalization 
of $S_{CS}$, while the second part is the spin-2 version of section 
\ref{sec:dim-1-conformal} where we compute the whole correlators.
  
First, to determine $c$ in $S_{\rm CS}$, let us look at the divergence of 
$\langle L_{\mu\nu} \, L_{\rho\sigma} \rangle$.    
From \eq{spin-2-LL-RR}, we have
\beq
  q^\mu \langle L_{\mu\nu} \, L_{\rho\sigma} \rangle
  = \frac{iN_c}{48\pi} \left[ A_{\nu\rho\sigma} + B_{\nu\rho\sigma} 
                             + C_{\nu\rho\sigma} \right]  \>,  
\eql{spin2-qLL}
\eeq
where
\beq
  A_{\nu\rho\sigma}
  &=& 2\tilde{q}_\nu \tilde{q}_\rho \tilde{q}_\sigma  \>,\\
  B_{\nu\rho\sigma}
  &=& 2q_\nu q_\rho q_\sigma - q^2 \eta_{\rho\sigma} q_\nu
      - q^2 \eta_{\nu\rho} q_\sigma - q^2 \eta_{\nu\sigma} q_\rho  \>,\\ 
  C_{\nu\rho\sigma}
  &=& 2q_\nu q_\rho q_\sigma 
      + q^2 \eta_{\rho\sigma} \tilde{q}_\nu 
      + q_\nu (\tilde{q}_\rho q_\sigma + q_\rho \tilde{q}_\sigma)  \>.
\eeq
Here, the $B$ and $C$ terms are actually not interesting, since they can be 
completely reproduced by just adding {\it local} terms at 
the $z=0$ boundary.
Specifically, the $B$ term is reproduced by adding 
\beq
  \Delta{\cal S}_{z=0}^{(B)} 
   &=& -\frac{N_c}{48\pi} \int\! d^2x \, \bar{h}^{\mu\nu} (-q) 
        \biggl[( \eta_{\mu\nu} q_\rho q_\sigma 
                         +\eta_{\rho\sigma} q_\mu q_\nu)  \nn\\
                           && \hspace{10EM}
               -\frac{q^2}{2} ( \eta_{\mu\rho} \eta_{\nu\sigma}
                               +\eta_{\nu\rho} \eta_{\mu\sigma}
                               +3\eta_{\mu\nu} \eta_{\rho\sigma})  
        \biggr] \bar{h}^{\rho\sigma} (q)  \>, 
\eql{spin2-local-B}
\eeq
while the $C$ term by
\beq
  \Delta{\cal S}_{z=0}^{(C)}  
  = -\frac{N_c}{24\pi} \int\! \frac{d^2q}{(2\pi)^2} \, \bar{h}^{\mu\nu}(-q) 
        \bigl[ \eta_{\mu\nu} q_\rho^L q_\sigma^L + \eta_{\rho\sigma} q^L_\mu q^L_\nu
        \bigr] \bar{h}^{\rho\sigma}(q)  \>.
\eql{spin2-local-C}
\eeq
Repeating the same exercise for $\langle R_{\mu\nu} \, R_{\rho\sigma} \rangle$
leads the same results except that all $q^L$ are replaced by $q^R$.
Since they are local, they have no effect on the physics.
In the following discussions, we will simply ignore them (and the corresponding
$B$ and $C$ terms in \eq{spin2-qLL}).

It thus all comes down to getting the $A$ term in \eq{spin2-qLL}.
In terms of the source $\ell_{\mu\nu}(x)$ of $L_{\mu\nu}(x)$, 
it implies that the generating functional $W[\ell]$ should transform under 
$\ell_{\mu\nu} \to \ell_{\mu\nu}(x) + \del_\mu \tilde{\xi}_\nu 
+ \del_\nu \tilde{\xi}_\mu$ as
\beq
  W \too W - \frac{iN_c}{24\pi} \int\! \frac{d^2q}{(2\pi)^2} \,
             \tilde{\xi}^\nu(-q) \, A_{\nu\rho\sigma}
             \, \ell^{\rho\sigma} (q)  \>.    
\eeq
Since $A_{\nu\rho\sigma}$ is `parity odd' (i.e.~it contains an odd number of
$\epsilon$ tensors), it must come from varying $S_{\rm CS}$.  Indeed,
comparing this to \eq{CS-boundary} with $\bar{h}_{\mu\nu} = \ell_{\mu\nu}$, 
we see that this can be exactly reproduced
by the gravitational Chern-Simons term \eq{S_CS} if we choose
\beq
  c = \frac{N_c}{6\pi}  \>.
\eeq
Repeating the same exercise for $\langle R_{\mu\nu} \, R_{\rho\sigma} \rangle$
gives $c = -N_c/6\pi$ instead.
   
Now that the divergence of $\langle L_{\mu\nu} \, L_{\rho\sigma} \rangle$ is
completely reproduced, our next task is to calculate the correlator itself.
It is convenient to parameterize $h_{\mu\nu}$ as%
\footnote{Hereafter, we will drop the tildes of $\tilde{h}_{\mu\nu}$ and 
$\tilde{h}$ to avoid notational clutter.}
\beq
  h_{\mu\nu} 
  = \frac{q_{\mu} q_{\nu}}{q^2} \phi 
   +\frac{\eta_{\mu \nu}}{2} (h - \phi) 
   +\frac{q_{\mu} \tilde{q}_{\nu} + \tilde{q}_{\mu} q_{\nu}}{2 q^2} \chi  \>, 
\eeq
where $\tilde{q}_\mu \equiv \epsilon_{\mu\nu} q^\nu$.
An advantage of this decomposition is that it `diagonalizes' \eq{S_EH}:
\beq
  {\cal S}_{\rm EH} 
  = \frac{M_*}{8} \int\! \frac{d^2q}{(2\pi)^2}\frac{dz}{z} 
      (-\phi' \phi' + h' h' + \chi' \chi')  \>. 
\eql{S_EH-simplified}
\eeq
Note that there is no $q^2$ appearing here, i.e.~the 3d gravity has no
propagating degrees of freedom.
On the other hand, the Chern-Simons term \eq{S_CS} mixes $h$, $\phi$, and 
$\chi$ and introduces $q^2$ dependencies:  
\beq
  {\cal S}_{\rm CS} 
  = \frac{c}{4} \int\! \frac{d^2q}{(2\pi)^2} \, dz \left[ 
      - \frac{q^2}{2} (h-\phi) \chi' 
      +\frac{q^2}{2} (h'-\phi') \chi 
      +\chi' \phi'' - \chi''\phi' \right]  \>. 
\eql{S_CS-simplified}
\eeq
The $h$-$\phi$-$\chi$ variables are also convenient for analyzing gauge
transformation properties.  In terms of the longitudinal and transverse
components of $\xi_\mu(q)$ defined as
\begin{eqnarray} 
  \tilde{\xi}_\mu(q) = \frac{iq_\mu}{2q^2} \, \xi_\parallel (q) 
                      +\frac{i\tilde{q}_\mu}{2q^2} \, \xi_\perp (q)  \>,
\end{eqnarray}
the residual gauge transformation \eq{grav-gauge-trans-residual} can be 
written as
\beq
  \phi(q,z) &\too& \phi(q,z) + \xi_\parallel (q)  \>,\\
  h(q,z)    &\too& h(q,z)    + \xi_\parallel (q)  \>,\\
  \chi(q,z) &\too& \chi(q,z) + \xi_\perp (q)  \>.
\eql{residual_gauge_trans}
\eeq
The advantage of this notation is that we immediately see that $h-\phi$ is gauge
invariant.

Now, following our gauge choice \eq{grav-gauge}, the constraint
equations are (see Appendix \ref{app:spin2-const-eq} for the derivation):
\beq
  \frac{h'}{z} - \frac{q^2}{2}(h - \phi) + \frac{c}{M_*} q^2 \, z\chi' 
    &=& 0  \>,\eql{const_eq_1}\\
  \frac{h'}{z} - \frac{\phi'}{z} - \frac{2c}{M_*}\, \chi''
    &=& 0  \>,\eql{const_eq_2}\\
  \frac{\chi'}{z} + \frac{c}{M_*} \bigl[ 2\phi'' - q^2 (h-\phi) \bigr]
    &=& 0  \>.\eql{const_eq_3}  
\eeq
One may also derive the equations of motion by varying the action 
${\cal S}_{\rm EH} + {\cal S}_{\rm CS}$ with respect to $h_{\mu\nu}$.
However, those equations of motion are redundant---they all can be derived
from the constraint equations \eq{const_eq_1}-\eq{const_eq_2}.
  
Now, we can use the constraint equations \eq{const_eq_1}-\eq{const_eq_3} 
to simplify the action ${\cal S}_{\rm EH} + {\cal S}_{\rm CS}$ and write it 
as boundary terms:
\beq
  {\cal S}_{\rm EH} + {\cal S}_{\rm CS} 
  = \int\! \frac{d^2q}{(2\pi)^2} 
      \left[ \frac{M_* q^2}{16} (h - \phi)^2
            +\frac{c q^2}{8} (h - \phi) \chi 
            -\frac{c q^2}{8} (h - \phi) \, z\chi' 
      \right]_{z=\epsilon}^{z=\infty}  \>.  
\eql{eff_action}
\eeq
Next, notice that the constraint equations 
(\ref{eq:const_eq_1})-(\ref{eq:const_eq_3}) imply
\beq
  z^2\chi''' + z\chi'' + \left( q^2z^2 - \alpha^2 \right) \chi' = 0
    \>, 
\eeq
where $\alpha \equiv M_*/2c$, and
\begin{eqnarray}
  h(z) - \phi(z) 
    = \bar{h} - \bar{\phi} +\frac{1}{\alpha} [z\chi'(z) - \chi(z)]
      -\frac{1}{\alpha} [\epsilon\chi'(\epsilon) - \bar{\chi}]  \>,
\end{eqnarray}
where the barred fields denote the corresponding 2d sources at the $z=\epsilon$
boundary, i.e., $\bar{h}(q) \equiv h(q,\epsilon)$, etc.
In the AdS/CFT correspondence, the 2d sources are located only at the 
$z=\epsilon$ boundary, so both $h-\phi$ and $\chi'$ must vanish as 
$z \to \infty$ (for Euclidean momenta $q^2 \equiv -Q^2 < 0$) so that 
the action \eq{eff_action} only gets contributions from the $z=\epsilon$ end.  
From the above expression of $h-\phi$, we see that $h-\phi$ can vanish only
if $\chi'$ is exponentially damped (hence $\chi$ approaches a constant) as 
$z \to \infty$, that is, only if $\chi'$ is proportional to 
$K_{|\alpha|} (Qz)$, without $I_{|\alpha|} (Qz)$ component.  Furthermore, 
since $\chi'$ is invariant under the residual gauge transformation 
\eq{residual_gauge_trans}, the proportionality factor can only depend on
$\bar{h} - \bar{\phi}$, but not on $\bar{\chi}$.  Therefore, we have   
\beq
  \chi' (z) = A \, (\bar{h} - \bar{\phi}) \, K_{|\alpha|} (Qz)  \>, 
\eeq
where $A$ is a numerical constant to be determined below.
Integrating this then gives
\beq
  \chi(z) = \bar{\chi} 
           + A \, (\bar{h} - \bar{\phi}) 
               \int_\epsilon^z\!\! K_{|\alpha|} (Qz') \, dz' \>. 
\eeq
So requiring that $h(z)-\phi(z)$ vanish at $z = \infty$, we have
\beq
  0 &=& \bar{h} - \bar{\phi} - \frac{1}{\alpha} \chi(\infty)
        -\frac{1}{\alpha} [\epsilon\chi'(\epsilon) - \bar{\chi}]
        \nonumber\\
    &=& (\bar{h} - \bar{\phi}) 
          \left[ 1 - \frac{A}{\alpha} 
                       \left( \int_\epsilon^\infty\!\! K_{|\alpha|} (Qz') \, dz'
                             + \epsilon \, K_{|\alpha|} (Q\epsilon) 
                       \right) 
          \right]  \>.
\eeq
This determines $A$, and we find
\beq
  \chi'(z) 
  = (\bar{h} - \bar{\phi}) \, 
      \frac{\alpha\, K_{|\alpha|} (Qz)}
           {\int_\epsilon^\infty\! K_{|\alpha|} (Qz') \, dz' 
            + \epsilon \, K_{|\alpha|} (Q\epsilon)}  \>.
\eql{chiprime}
\eeq
For $|\alpha| \geq 1$, this implies
\beq
  \lim_{\epsilon \to 0} \epsilon \chi'(\epsilon) 
  = {\rm sgn}(c) \, (|\alpha| - 1) (\bar{h} - \bar{\phi})  \>,
\eeq
where ${\rm sgn}(c)$ is the sign of $c$, namely, $+1$ for 
$h_{MN} = {\cal L}_{MN}$ while $-1$ for 
$h_{MN} = {\cal R}_{MN}$.  
Then, putting this into \eq{eff_action}, we obtain
\beq
  {\cal S}_{\rm EH} + {\cal S}_{\rm CS} 
  &=& \int\!\! \frac{d^2q}{(2\pi)^2} 
        \left[ -\frac{M_* q^2}{16} (\bar{h} - \bar{\phi})^2
               -\frac{c q^2}{8} (\bar{h} - \bar{\phi}) \bar{\chi} 
               +\frac{|c| (|\alpha|-1) q^2}{8} (\bar{h} - \bar{\phi})^2 
        \right]  \nonumber\\
  &=& \int\!\! \frac{d^2q}{(2\pi)^2}
        \left[ -\frac{c q^2}{8} (\bar{h} - \bar{\phi}) \bar{\chi} 
               -\frac{|c| q^2}{8} (\bar{h} - \bar{\phi})^2
        \right]  \>,
\eql{eff_action_final}
\eeq
where we see that the $M_*$ term has completely cancelled out since 
$\alpha = M_*/2c$.  This is exactly analogous to what has happened to the 
${\cal L}_M$-${\cal R}_M$ sector in section \ref{sec:dim-1-conformal}
where the result became completely independent of the value of $g_3$.

To check that the above result agrees with the 2d result \eq{spin-2-LL-RR},
let us translate the result \eq{eff_action_final} back to the original 
$h_{\mu\nu}$ variable, note that
\beq
  \bar{h} - \bar{\phi}
    = -2\frac{\tilde{q}_\mu \tilde{q}_\nu}{q^2} \bar{h}^{\mu\nu}
      \quad, \qquad
  \bar{\chi}
    = -2\frac{q_\mu \tilde{q}_\nu}{q^2} \bar{h}^{\mu\nu}
          \>.
\eeq
Then, \eq{eff_action_final} becomes 
\beq
  {\cal S}_{\rm EH} + {\cal S}_{\rm CS} 
  = -\int\! \frac{d^2q}{(2\pi)^2} \bar{h}^{\mu\nu}(-q)
       \biggl[ \frac{c}{4}  
               \frac{ \tilde{q}_\mu \tilde{q}_\nu q_\rho \tilde{q}_\sigma
                     +q_\mu \tilde{q}_\nu \tilde{q}_\rho \tilde{q}_\sigma}{q^2}      
              +\frac{|c|}{2} 
               \frac{\tilde{q}_\mu \tilde{q}_\nu 
                     \tilde{q}_\rho \tilde{q}_\sigma}{q^2}
       \biggr] \bar{h}^{\rho\sigma}(q) \>.       
\eeq
Let us check this for the $LL$ correlator (i.e.~$c= + N_c/6\pi$, and 
$\bar{h}_{\mu\nu} = \ell_{\mu\nu}$).  Then, this formula gives
\beq
  \langle L_{\mu\nu} \, L_{\rho\sigma} \rangle (q)
  &=& \frac{iN_c}{24\pi} \frac{ 4 q_\mu q_\nu q_\rho q_\sigma 
                               +q_\mu q_\nu ( q_\rho \tilde{q}_\sigma 
                                             +\tilde{q}_\rho q_\sigma) 
                               +(\tilde{q}_\mu q_\nu + q_\mu \tilde{q}_\nu) 
                                q_\rho q_\sigma}{q^2}  \nn\\
  &&  + (\text{local terms})  \>.
\eeq
Notice that the 2d formula \eq{spin-2-LL-RR} has exactly 
the same nonlocal piece.

Finally, let us comment briefly on what happens if $|\alpha|<1$.  
In this case, the $\epsilon \to 0$ limit converges in the denominator of 
\eq{chiprime}, so $\chi'(z)$ is just $K_{|\alpha|}(Qz)$ times an 
$\epsilon$-independent factor.  Then, 
$\epsilon\chi'(\epsilon) \propto \epsilon^{1-|\alpha|} \to 0$ as 
$\epsilon \to 0$, therefore the last term in \eq{eff_action_final} vanishes.
In this case, the 3d calculations would agree with the 2d results only if
$M_* = 2|c|$ which, however, is outside the range $|\alpha| < 1$.  Therefore,
the $|\alpha| < 1$ case would lead to wrong correlators.  On the other hand, 
the correlators from the 3d side are correct for any $|\alpha| \geq 1$, as we 
have seen above.

\subsubsection{Higher-Spin Operators}
\label{sec:higher-spin}

The general features common to the correlators between primary operators 
with spin $>2$ (i.e.~$k>2$ in \eq{conformal-O_k-O_l}) are all 
already present in the spin-2 case discussed in section \ref{sec:dim-2}.
Here we just summarize those features.  First, just like the case with any $k$, 
there are two bulk fields ${\cal L}_{M_1 \cdots M_k}$ and
${\cal R}_{M_1 \cdots M_k}$ (all the indices being symmetrized)
corresponding to the left- and right-moving sectors in 2d.  
As usual, we only focus on the two-point correlators, so we are only concerned with the quadratic part of the action for ${\cal L}_{M_1 \cdots M_k}$ and 
${\cal R}_{M_1 \cdots M_k}$.
In this case, the `kinetic term' (the analog of 
${\cal F}_{MN} {\cal F}^{MN}$ of the $k=1$ case or 
${\cal S}_{\rm EH}$ of the $k=2$ case) is constrained by the generalization 
of the gauge transformation \eq{grav-gauge-trans-full} where 
the gauge-transformation parameter $\xi_A$
is replaced by a traceless, totally-symmetric rank-$(k-1)$ tensor 
$\xi_{M_1 \cdots M_{k-1}}$.  (A trace{\it ful} component would be the
gauge-transformation parameter for a field with lower $k$.)   
They also have the analog of the Chern-Simons term
${\cal S}_{\rm CS}$.  While the `kinetic' term is identical for the left
and right sectors, their `Chern-Simons' terms differ by a sign.  This aspect
is common to all $k$.

Now, one of the properties shared by all $k \geq 2$ cases (but not by $k=1$) 
is that the equations of motion are all redundant and can be derived from the
constraint equations.  (We have seen this in the spin-2 case, while in the
spin-1 case there is one real equation of motion \eq{L-EOM-conformal}.)  
This can be understood by a simple counting.  For example, for ${\cal L}_{LMN}$, 
we begin with $3\cdot 4 \cdot 5/3! = 10$ components, but by using the 
$3 \cdot 4/2! -1 = 5$ gauge parameters, we can set 5 components to zero, 
so there are 5 constraint equations (the analogs of 
\eq{const_eq_1}-\eq{const_eq_3}).
The remaining $5$ components of ${\cal L}_{LMN}$ have 5 equations of motion, but these must be all redundant since we already have the 5 constraint
equations and the constraint equations are lower order in derivatives.  
Therefore we have only constraints and no real equations of motion.
However, this does not mean that the equations are trivial.
As we have seen in the spin-2 case, the Chern-Simons term can make the 
constraint equations depend on $q^2$, thus effectively introducing 
propagation.  Note, however, that the detailed form of the propagating modes 
did not play a significant role in reproducing the correlation functions.

Next, the structure of the `Chern-Simons' term is the following.  The (quadratic
part of) `Chern-Simons' term should {\it contain} the structure 
$\epsilon \, {\cal L} \, \del \, {\cal L}$, i.e., one $\epsilon$ tensor 
(3 upper indices), two ${\cal L}$ fields ($2k$ lower indices) with one
derivative in between (1 lower indices).  But there are still $2k-2$ lower indices yet to be contracted.  Furthermore, it needs to have the right scaling
property under $x^M \to \lambda x^M$ to be consistent with the AdS$_3$
isometry.  Since the kinetic term has the form 
$\int\! d^3x \sqrt{g} \, (g^{-1})^{k+1} \nabla {\cal L} \, \nabla {\cal L}$
where $g^{-1}$ denote the inverse metric, ${\cal L}$ must scale as 
${\cal L} \to \lambda^{-k} {\cal L}$.  Thus, the object that 
gets contracted with the $2k-2$ lower indices in the Chern-Simons term
must scale as $\lambda^{2k-2}$.  The only way to do this is to have additional
$2k-2$ derivatives and $2k-2$ inverse metrics.  Hence, schematically, 
(the quadratic part of) the `Chern-Simons' term has 
the form $\int\! d^3x\, \epsilon \, {\cal L} \, (g^{-1})^{2k-2} 
\nabla^{2k-1} {\cal L}$ where the indices are contracted in various ways such 
that the whole thing becomes gauge invariant (up to a surface term) 
under the gauge transformation mentioned above.  
Note that this form agrees with what we have explicitly written down for the 
$k=1$ and $k=2$ cases.  

Finally, we expect that, like in the $k=1,2$ cases, once we fix the
coefficients of the `Chern-Simons' terms by matching the divergences of
the current-current correlators, the whole correlators (in the conformal limit)
should be automatically reproduced regardless of the coefficients of the
`kinetic' terms.    
However, there is a notable difference between the $k=1$ case and all $k\geq 2$
cases.  The $k=1$ Chern-Simons is special because it contains no metric, so it
is insensitive to a deformation of the bulk geometry.  This was the essential 
reason why the $k=1$ correlators in the conformal limit is actually exact to
all orders in $\Lambda$.  On the other hand, since all $k\geq 2$ Chern-Simons
terms depend on the metric, so the $k\geq 2$ correlators should receive 
corrections depending on $\Lambda$, which is in accord with the 2d results.

\subsubsection{The $U(1)_A$-Charged Sector}
\label{sec:scalar-conformal}

This sector only contains one operator $X$.  Since $X$ is 
a dimension-one operator, the corresponding bulk scalar field 
${\mathcal X}$ has mass-squared $-1$.  Therefore, the quadratic part of 
the scalar-sector action (with the short-distance cutoff $\epsilon$) 
is given by
\beq
  {\cal S}_X
   = \int\! d^2x \int_\epsilon^\infty \!\! dz  
     \biggl[ \frac{1}{z} (\del_M {\cal X}^\dagger) \del^M\! {\cal X}
            + \frac{1}{z^3} {\cal X}^\dagger {\cal X} \biggr]  \>.
\eql{conformal-X-action}
\eeq
For 2d momentum $q$, the equation of motion from this action reads
\beq
  z^2 {\cal X}'' - z{\cal X}' - (Q^2 z^2 - 1) {\cal X} = 0  \>, 
\eql{X-EOM-conformal}
\eeq
where $Q^2 \equiv -q^2$.  The solution satisfying the boundary condition 
$\lim_{z \to \infty} {\cal X} \to 0$ is 
\beq
  {\cal X}(q,z) 
    = Z_X^{-1/2} \, \frac{z \, K_0(Qz)}{\epsilon \, K_0(Q\epsilon)} \, 
      J_X(q)  \>,
\eql{conformal-X-solution}
\eeq
where $J_X (q,\epsilon)$ is the (renormalized) source for $X(q)$, with 
the wavefunction renormalization $Z_X$.  

Since we are in the conformal limit (i.e.~$\Lambda \to 0$ and $m_q \to 0$), 
it is diagrammatically straightforward to compute 
$\langle X^\dagger X\rangle$ in the 2d side, which gives 
\beq
  \langle X^\dagger X \rangle (q) 
    = \frac{i N_c}{\pi} \log Q + \cdots  \>,
\eeq
where the $\cdots$ refers to a scheme-dependent local piece.  
On the other hand, the effective action obtained by plugging 
\eq{conformal-X-solution} into \eq{conformal-X-action} yields
\beq
  \langle X^\dagger X \rangle (q)
   \too -\frac{i}{\epsilon^2 Z_X} \frac{1}{\log Q\epsilon} 
        + \cdots \>,
\eeq
where $\cdots$ denotes terms which are local or higher-order in 
$\epsilon$.  To subtract the $\epsilon$ dependence, we have 
to introduce a fixed (but arbitrary) renormalization scale 
$\mu \ll \epsilon^{-1}$.  (This dependence on $\mu$ precisely reflects 
the scheme dependence of the finite term in the 2d side.)  Then, we rewrite 
$\log Q\epsilon$ as $\log Q\epsilon = \log\mu\epsilon + \log(Q/\mu)$,  
and the above expression becomes 
\beq
   \langle X^\dagger X \rangle (q)
  \too \frac{i}{\epsilon^2 Z_X} 
       \frac{\log(Q/\mu)}{(\log\mu\epsilon)^2} + \cdots  \>.
\eeq
Hence, $Z_X^{-1/2}$ must be proportional to $\epsilon \log\mu\epsilon$ in order for 
the $\epsilon \to 0$ limit to be finite.  Matching the coefficients of $\log Q$,
we determine the wavefunction renormalization: 
\beq
  Z_X^{-1/2} = \sqrt{\frac{N_c}{\pi}} \, \epsilon \log\mu\epsilon  \>.  
\eql{Z_X}
\eeq
Thus we have exactly reproduced $\langle X^\dagger X \rangle$ in the conformal
limit.

\subsection{Conformal Symmetry Breaking at $O(\Lambda)$}
\label{sec:conf-break}

As we pointed out in section \ref{sec:first-order-mixing}, the only 
nonzero correlators at $O(\Lambda)$ are 
$\langle X L_{\mu_1 \cdots \mu_k} \rangle$ and 
$\langle X R_{\mu_1 \cdots \mu_k} \rangle$ (and their Hermitian conjugates).
This means that at ${\cal O}(\Lambda)$, the only effect of 
the breaking of conformal invariance is the `apparent' chiral symmetry breaking
discussed in section \ref{sec:chiral-sym-break}.  The corresponding 3d analyses
are quite analytically tractable because the geometry can be still taken to be
AdS$_3$; note that a deviation from AdS$_3$ would lead to 
$\langle T_\mu^\mu \rangle \neq 0$ for the 2d stress-tensor, but 
from dimensional analysis this must be proportional to $\Lambda^2$.  
Therefore, for ${\cal O}(\Lambda)$ analyses, there is no need to
worry about backreaction to the geometry.  Therefore, we begin with 
the ${\cal O}(\Lambda)$ case (which includes some exact results, as we
advertised earlier),  then move on to analyses at ${\cal O}(\Lambda^2)$.

First, notice that the only source of $O(\Lambda)$ effects is $X$ 
(see section \ref{sec:first-order-mixing}).  Hence, in the 3d side, 
we must be able to describe all $O(\Lambda)$ effects in terms of 
$\langle {\cal X} \rangle$.
In particular, as we already pointed out, the geometry can be taken to be just
AdS$_3$.
 
For definiteness and simplicity, let us just focus on the $PL_\mu$ and 
$PR_\mu$ correlators, \eq{PL-massless} and \eq{PR-massless}. 
$P$ also mixes with $L_{\mu_1 \cdots \mu_k}$ and $R_{\mu_1 \cdots \mu_k}$ 
with $k=3,5,\cdots$, but this could affect $\langle P L_\mu \rangle$ and 
$\langle P R_\mu \rangle$ only at $O(\Lambda^2)$ or higher.  Actually, 
since \eq{PL-massless} and \eq{PR-massless} are exact, there are 
no higher-order corrections to them; we will see below from a 3d viewpoint 
why they are exact.

As we discussed in section \ref{sec:chiral-sym-break}, 
the $O(\Lambda)$ effects in the correlators \eq{PL-massless} 
and \eq{PR-massless} describe (apparent) chiral symmetry breaking.
Therefore, the corresponding 3d physics must be spontaneous breaking of 
$U(1)_A$ by nonzero $\langle {\cal X} \rangle$, giving a mass to 
${\cal A}_M = {\cal L}_M - {\cal R}_M$ (but not to 
${\cal V}_M = {\cal L}_M + {\cal R}_M$).  We parameterize ${\cal X}$ as  
\beq
  {\cal X} = \left( \langle {\cal X} \rangle
             + \frac{{\cal H}}{\sqrt2} \right) e^{i{\cal \widetilde{\cal P}}}
             \>,
\eeq
where ${\cal H}(x,z)$ is a real scalar field with $\langle {\cal H} \rangle =0$,
while $\widetilde{{\cal P}}$ is a Goldstone field which shifts as 
$\widetilde{{\cal P}} \to \widetilde{{\cal P}} - \alpha$ under the $U(1)_A$
gauge transformation ${\cal A}_M \to {\cal A}_M + \del_{M} \alpha$.
Since $X=(S+iP)/\sqrt2$, the real scalar field ${\cal P}$ that corresponds to the 
2d operator $P$ is given by
\beq
  \widetilde{{\cal P}} = \frac{\cal P}{\sqrt{2} \langle {\cal X} \rangle}  \>.
\eql{Ptilde-def}
\eeq
Now, since ${\cal H}$ and ${\cal P}$ do not couple to each other at the
quadratic order, we can ignore ${\cal H}$ for the purpose of studying 
$\langle P L_\mu \rangle$ and $\langle P R_\mu \rangle$.  Then, 
the $U(1)_A$ gauge invariance tells us exactly how the actions \eq{S_LR} and 
\eq{conformal-X-action} must be combined:
\beq
  {\cal S}_{L,R,P} 
    = {\cal S}_{L,R}
     + \int\! d^2x \int_\epsilon^\infty\!\! dz \,
         \frac{\langle {\cal X} \rangle^2}{z} 
           (\del_M \widetilde{{\cal P}} + {\cal A}_M) 
           (\del^M \widetilde{{\cal P}} + {\cal A}^M)  \>.
\eql{LRX-action}
\eeq
What is $\langle {\cal X} \rangle$?  Note that if the geometry were exactly
AdS$_3$, the $X$ equation of motion \eq{X-EOM-conformal} would tell us that 
$\langle {\cal X} \rangle \propto \Lambda z$.  The mass of ${\cal A}_M$ would 
then be $\propto \Lambda z$, which would not be AdS$_3$ invariant.  Hence the
geometry cannot be exactly AdS$_3$, but, as we already mentioned, the 
deviation from AdS$_3$ is an $O(\Lambda^2)$ effect, so it is consistent to
say that background is  AdS$_3$ with 
$\langle {\cal X} \rangle \propto \Lambda z$ as long as we are only 
concerning $O(\Lambda)$ effects.  Therefore, we parameterize 
$\langle {\cal X} \rangle$ as
\beq 
  \langle {\cal X} \rangle = \kappa\Lambda z + O(\Lambda^2 z^2)  \>,    
\eql{X-VEV}
\eeq
and the determination of $\kappa$ does not get affected by higher order
effects.
 
We stick with the gauge choice ${\cal L}_3 = {\cal R}_3 = 0$, 
but now the constraint \eq{L-constraint-conformal} and its ${\cal R}$ 
counterpart are modified:
\beq
  \frac{1}{g_3^2} z {\cal L}'_\parallel + \frac{N_c}{2\pi} {\cal L}_\perp
   - \frac{2 \langle {\cal X} \rangle^2}{z} \widetilde{{\cal P}}'
   &=& 0  \>, \nn\\ 
  \frac{1}{g_3^2} z {\cal R}'_\parallel - \frac{N_c}{2\pi} {\cal R}_\perp
   + \frac{2 \langle {\cal X} \rangle^2}{z} \widetilde{{\cal P}}'
   &=& 0  \>. 
\eql{LR-constraint-Lambda}
\eeq
Then, the analog of the effective action \eq{S_LR-surface} is given by
\beq
  {\cal S}_{L,R,P}
   = \text{[r.h.s. of \eq{S_LR-surface}]}  
       + \int\! \frac{d^2q}{(2\pi)^2} 
         \frac{\langle {\cal X}\rangle^2}{\epsilon} \biggl[
           \widetilde{{\cal P}}(-q) \, \widetilde{{\cal P}}'(q) 
           + \frac{1}{Q^2} {\cal A}_\parallel (-q) \, \widetilde{{\cal P}}' (q)           
         \biggl]_{z=\epsilon}  \>.
\eql{S_LRP-surface}
\eeq
Now, note that the ${\cal A}_\parallel \widetilde{{\cal P}}'$ term above 
gives an $O(\Lambda)$ contribution to the $PL$ and $PR$ correlators.  
More explicitly, from \eq{Z_X}, \eq{Ptilde-def}, and \eq{X-VEV}, we get
\beq
  \frac{\langle {\cal X}\rangle^2}{\epsilon} \widetilde{{\cal P}}' (q, \epsilon)
  &=& \kappa \Lambda \sqrt{\frac{N_c}{2\pi}} 
      \frac{\log\mu\epsilon}{\log Q\epsilon} \, J_P(q) + O(\epsilon)  \nn\\
  &\too& \kappa \Lambda \sqrt{\frac{N_c}{2\pi}} \, J_P(q)  \>,         
\eql{XPprime}
\eeq
where $J_P(q)$ is the (renormalized) source for $P(q)$.  Note that this result
is actually exact, because corrections which are higher order in $\Lambda$ 
are necessarily accompanied by higher powers of $z$, hence will vanish when
the $\epsilon \to 0$ limit is taken.  This formula together with \eq{Ptilde-def}) tells us that 
the ${\cal A}_\parallel\widetilde{{\cal P}}'$ term in
\eq{S_LRP-surface} are $O(\Lambda)$, while the 
$\widetilde{{\cal P}}\widetilde{{\cal P}}'$ term is still purely $O(\Lambda^0)$,
which is consistent with our observation that the corrections to the $PP$ 
correlator begins at $O(\Lambda^2)$.

There are other places where $O(\Lambda)$ contributions appear;  
It is no longer true that in the r.h.s.~of \eq{S_LR-surface} we can replace
$\epsilon{\cal L}'_\perp$ and $\epsilon{\cal R}'_\perp$ with 
$-\nu {\cal L}_\perp$ and $-\nu {\cal R}_\perp$.  Now, ${\cal L}'_\perp$ and 
${\cal R}'_\perp$ contain an $O(\Lambda)$ piece. 
To see this, we must look at the equation of motion for ${\cal L}_\perp$ 
and ${\cal R}_\perp$: 
\beq
 \frac{1}{g_3^2} \bigl[ z(z {\cal L}_\perp')' - Q^2 z^2 {\cal L}_\perp \bigr]
   + \frac{N_c}{2\pi} z {\cal L}'_\parallel 
   - 2 \langle {\cal X} \rangle^2
       \bigl( {\cal L}_\perp - {\cal R}_\perp \bigr) &=& 0  \>, \nn\\
 \frac{1}{g_3^2} \bigl[ z(z {\cal R}_\perp')' - Q^2 z^2 {\cal R}_\perp \bigr]
   - \frac{N_c}{2\pi} z {\cal R}'_\parallel 
   + 2 \langle {\cal X} \rangle^2 
       \bigl( {\cal L}_\perp - {\cal R}_\perp \bigr) &=& 0  \>.
\eql{LR-EOM-Lambda}
\eeq
Combining these with \eq{LR-constraint-Lambda} and throwing away terms 
$O(\Lambda^2)$ or higher, we get
\beq
 \frac{1}{g_3^2} \bigl[ z(z {\cal L}_\perp')' 
                       - (Q^2 z^2 + \nu^2) {\cal L}_\perp \bigr]
 = -\frac{2\nu \langle {\cal X} \rangle^2}{z} 
      \widetilde{{\cal P}}'  \>, 
\eeq
where $\nu = g_3^2 N_c/(2\pi)$ as before, and the corresponding equation for
${\cal R}_\perp$ is identical.  Now, we write ${\cal L}_\perp$ as
${\cal L}_\perp^{(0)} + {\cal L}_\perp^{(1)}$ where ${\cal L}_\perp^{(0)}$
is the conformal solution \eq{L_perp-sol} and ${\cal L}_\perp^{(1)}$ is
the $O(\Lambda)$ perturbation.  Then, the perturbation satisfies
\beq
  z(z {\cal L}^{(1)\prime}_\perp)' - (Q^2 z^2 + \nu^2) {\cal L}^{(1)}_\perp 
    &=& -\frac{2\nu g_3^2 \langle {\cal X} \rangle^2}{z} 
         \widetilde{{\cal P}}^{(0)\prime}  \nn\\
    &=& - \nu g_3^2 \kappa \Lambda \sqrt{\frac{2 N_c}{\pi}} \, J_P(q) + O(z) 
        \>, 
\eeq
where the `source term' approaches a constant for small $z$, as seen
in the last line above.  Then, the small-$z$ behavior of 
the perturbation is
\beq
  {\cal L}^{(1)}_\perp
   = -\frac{g_3^2 \kappa \Lambda}{\nu} \sqrt{\frac{2 N_c}{\pi}} \, J_P(q) \,
      \left( \frac{\epsilon}{z} \right)^\nu + \cdots  \>,
\eeq
where the $\cdots$ refers to subleading terms for small $z$.  When we 
re-evaluate ${\cal L'}_\perp$ in \eq{S_LR-surface} by taking 
${\cal L}^{(1)}_\perp$ into account, we get a new term proportional to 
${\cal L}_\perp {\cal P}$, and repeating these steps for ${\cal R}_\perp$
gives the same coefficients for ${\cal R}_\perp {\cal P}$.
Putting all the pieces together, \eq{S_LRP-surface} becomes
\beq
  {\cal S}_{L,R,P}
   &=& \text{[r.h.s. of \eq{3d-LL-RR}]}  
       + \int\! \frac{d^2q}{(2\pi)^2} 
         \frac{\langle {\cal X}\rangle^2}{\epsilon} \biggl[
           \widetilde{{\cal P}}(-q) \, \widetilde{{\cal P}}'(q) 
         \biggl]_{z=\epsilon}  \nn\\
       &&
       + i\kappa\Lambda \sqrt{\frac{N_c}{2\pi}}
         \int\! \frac{d^2q}{(2\pi)^2} \frac{1}{Q^2} 
         \bigl\{ q_\mu^L {\cal L}^\mu (-q) - q_\mu^R {\cal R}^\mu (-q) \bigr\}
          J_P(q)  \>. 
\eql{3d-LRP}
\eeq
This exactly reproduces the 2d results \eq{PL-massless} and \eq{PR-massless}
if we choose
\beq
  \kappa = \sqrt{\frac{2\pi N_c}{3}}  \>.
\eeq

Looking back at the above calculation, we notice that the results are 
completely determined by the leading small-$z$ behavior of 
$\langle {\cal X} \rangle^2 \widetilde{{\cal P}}'$.  Since terms higher-order
in $\Lambda$ always come with higher-powers of $z$, the leading small-$z$
behavior of $\langle {\cal X} \rangle^2 \widetilde{{\cal P}}'$ 
calculated above will not get corrected.  Therefore, the formulas 
\eq{PL-massless} and \eq{PR-massless} are exact in the dual theory,
as they are in 2d!

\section{Mapping the 2d Theory to AdS$_3$}
\label{sec:transform}

Let us summarize what we have done so far.  We have constructed the
corresponding quadratic action for a certain bulk fields.  In the quadratic action,
the complexity of conformal symmetry breaking effects are encoded in mixings
of the bulk fields.  In principle, the mixings can be systematically identified 
by continuing what we did in section \ref{sec:dual} to include other fields,
order-by-order in $\Lambda z$.  However, this way of getting the 3d action---by
computing correlators and comparing them with the 2d results---seems quite `indirect'.  
In other words, on the one hand we have the 't Hooft equation,
which encodes all information about two-point correlators, while on the other
hand we are interested in the form of the (linearized) equations of motion for
the bulk fields, and in particular, the mixings.
However, to map one side to the other, we had to solve the equations and match
the solutions, which is an extra step.  It is much more
desirable to have a {\it direct} map from the 't Hooft equation to the 
equations of motion for the bulk fields.   

To this goal, we again follow our general philosophy and begin with the
conformal limit of the 't Hooft equation, and try to see if we can directly 
map it to an equation of motion in AdS$_3$.  But {\it which} equation of motion?
While the 't Hooft equation is a single equation, there are an infinite number
of equations of motion in the 3d side because there are infinite number of
fields.  To answer this question, recall that in the conformal
limit the $SS$ and $PP$ correlators are the only ones that {\it know} about
the nontrivial dynamics of the full model. 
The correlators among $U(1)_A$-neutral currents (such as $L_\mu$ and 
$R_{\mu\nu}$) all have just a $1/q^2$ pole without any other non-analytic
structure.  In other words, in taking the $\Lambda \to 0$ limit, all the poles
$1/(q^2 - m_n^2)$ have collapsed down to $1/q^2$.  This pole has completely lost
information about dynamics, since as seen from the 3d perspective, the 
residue of the pole is completely determined by the coefficient of the 
Chern-Simons term, i.e.~by the anomalies.
The scalar $S$ or pseudo-scalar $P$ two-point
functions, on the other hand, have logarithmic behavior at high energies.
These are obtained by summing over {\it all} the mesons, where the sum goes as
$\sum_n 1/(q^2- \Lambda^2 n) \sim \log(-q^2)$ (recall that 
$m_n^2 \simeq \pi^2\Lambda^2 n$ for $n \gg 1$.)  That is, the contributions 
from the highly excited states are crucial for obtaining the logarithmic
behavior expected from the asymptotic freedom.  We therefore cannot
simply take $\Lambda$ to zero and collapse all $m_n$ to zero, but rather we need 
to take $\Lambda \rightarrow 0$ and $n \rightarrow \infty$ with 
$m_n^2 \sim \pi^2 \Lambda^2 n$ fixed.
We thus expect that if we take this scale invariant limit of the 't Hooft
equation for the parton wave function $\phi_n(x)$, it should be related to the
AdS equation of motion for the fields dual to operators $S$ and $P$.

This limit, which zooms in to the large-$n$ mesons and makes the 
scale invariance of the 't Hooft equation manifest, 
was first derived in \cite{Rich} in the context of 
analyzing the behavior of $\phi_n(x)$ near the `turning points' in the 
semi-classical approximation.
First, let us 
rescale the $x$-variable as $x \to \Lambda^2 x$ (followed by
the redefinition of $\phi_n$ as $\phi(\Lambda^2 x) \to \phi_n(x)$).
The 't Hooft equation \eq{tHooftEq} then reads
\beq
   \frac{\widetilde{m}_q^2 -1}{x (1 - \Lambda^2 x)} \, \phi_n (x) 
   -\hat{{\rm P}} \!\!\int_0^{1/\Lambda^2}\!\! 
      \frac{\phi_n (y)}{(x - y)^2} \, dy 
  = m_n^2 \, \phi_n (x)  \>,
\eeq
where $\widetilde{m}_q \equiv m_q/\Lambda$.
We now take the limit $\Lambda \to 0$ and $n \to \infty$ with $m_n^2 \equiv m^2$
fixed (and also $m_q \to 0$ with $\widetilde{m}_q$ fixed) 
to obtain
\beq                                                                                                     
 (\hat{T} \!*\! \phi)(m^2 x)
 \equiv \frac{\widetilde{m}_q^2-1}{x} \,\phi(m^2 x) 
       -\hat{{\rm P}} \!\!\int_0^{\infty} \! 
          \frac{\phi(m^2 y)}{(x-y)^2} \, dy = m^2 \phi(m^2 x)  \>,
\eql{tHooft_scale_inv}
\eeq
where we have written $\lim_{n \to \infty} \phi_n(x)$ as $\phi(m^2x)$ to 
make it explicit that $\phi$ only depends on the combination $m^2x$.
Now it is obvious that the equation has an invariance under 
$x \to \lambda x, m^2 \to m^2/\lambda$ with any positive constant $\lambda$.  
(Note that $m^2$ is now a continuous eigenvalue.)
Hence, in principle the equation \eq{tHooft_scale_inv} has all the necessary
ingredients 
to describe the conformal limit of the 't Hooft model, 
as we have discussed above. 
However, the full conformal symmetry, which is more than just scale invariance,
is not manifest in \eq{tHooft_scale_inv}, although it should be so secretly.

To reveal the hidden conformal invariance of \eq{tHooft_scale_inv}, note
the following identity:
\beq
   \int_0^{\infty} \! \frac{dx}{x} \, 
                \sin\!\left( \frac{\pi z^2}{4x} \right) \, 
                \cos\!\left( \frac{m^2 x}{\pi} \right) 
  = \frac{\pi}{2} \, J_0(mz) \>.                      
\eeq
Recalling the approximate form of the 't Hooft wavefunction \eq{approx-soln},
this suggests that we should consider the following transform of the 
$\phi(m^2 x)$wave function:
\beq
  \widetilde{\phi}(z) 
  = \int_0^{\infty} \!\! dx \,
      \del_z \phi\!\left( \frac{\pi^2 z^2}{4 x} \right) \, \phi (m^2 x) \>.
\eql{thetransform}
\eeq
Then, the above identity says that $\widetilde{\phi}(z) \propto z J_0(mz)$,
which is of course a solution of the equation of motion \eq{X-EOM-conformal} 
for the bulk scalar ${\cal X}$!  Being purely $J_0$ without a $Y_0$ component, 
it even satisfies the right boundary condition 
($\lim_{z \to 0} \widetilde{\phi}(z) \to 0$) to be a KK mode.%
\footnote{Strictly speaking, we do not have `KK modes' in the exact AdS$_3$ 
limit, but one should imagine that the geometry deviates from AdS$_3$ at 
large $z$ corresponding to the breaking of conformal symmetry in the 2d side.
Then our discussions here are valid for the small-$z$ behavior of 
$\widetilde{\phi}(z)$.}

Our goal is, however, to map equations to equations, rather than solutions to solutions.  
Thus, let us check that the above transform maps the
scale-invariant limit of the 't Hooft equation \eq{tHooft_scale_inv} to 
a bulk equation of motion in AdS$_3$.
First, notice
that from \eq{tHooft_scale_inv}, one can show that the operator $\hat{T}$ 
has the property that
\beq
  \int_0^\infty\! dx \, f\!\left(\frac{u^2}{x}\right) \,
    (\hat{T}\!*\!g)(m^2x)
  = \int_0^\infty\!\! dx \, g\!\left(\frac{m^2}{x}\right) \,
    (\hat{T}\!*\!f)(u^2x)  \,
\eeq
for arbitrary functions $f$ and $g$.
Applying this to the case $f(u^2/x)= \phi(\pi^2 z^2/4x)$, 
we obtain
\beq
  \int_0^{\infty} \!\! dx \, 
    \del_z \phi \!\left( \frac{\pi^2 z^2}{4x} \right) \, 
    (\hat{T} \!*\! g)(m^2x)
 = \int_0^{\infty} \!\frac{dy}{y^2} \, 
     \del_z \!\!\left[ \frac{\pi^2 z^2}{4} 
                     \phi\!\left( \frac{\pi^2 z^2}{4y} \right) 
              \right] \, g(m^2 y)  \>,
\eeq
for any $g(m^2 x)$.
In the limit that $\widetilde{m}_q \to 0$, we have 
$\del_z \phi (\frac{\pi^2 z^2}{4x}) \simeq -\frac{\pi z}{\sqrt{2}x} 
\sin(\pi z^2/4x)$, and thus 
\beq 
  \frac{1}{y^2} \, \del_z\!\left( \frac{\pi^2 z^2}{4} \phi \right) 
 = -\left[ z \del_z (z^{-1} \del_z) + \frac{1}{z^2} \right] \!
    \del_z \phi  \>.
\eeq
Finally, letting $g=\phi$, we find that $\widetilde{\phi}(z)$ 
obeys the equation
\beq
  -\left[z \del_z (z^{-1} \del_z ) + \frac{1}{z^2}\right] \!\widetilde{\phi}(z) 
  = m^2 \widetilde{\phi}(z) \>,
\eeq
which is the appropriate wave equation in AdS$_3$ for a KK-mode of a scalar
field dual to a dimension one operator, i.e.~$X$.  Since it is mapped
to an AdS$_3$ invariant equation, the scale-invariant 't Hooft equation 
\eq{tHooft_scale_inv} is indeed fully conformally invariant.  The transform
\eq{thetransform} also shows an explicit connection between parton-$x$ and 
the radial coordinate $z$ of AdS$_3$.

In addition, note that the transform provides an explicit check of 
the AdS/CFT prescription.  Namely, consider the following kernel
\beq
  G_0(q^2,x) \equiv 
    m_q \!\sum_{n=0,2,4,\cdots}\! 
    \frac{\phi_n(x)}{q^2-m_n^2} \int_0^1\! \frac{dy}{y} \, \phi_n(y)  \>.
\eeq
(Here $\hat{T}$ refers to the exact 't Hooft operator rather than the
scale-invariant one \eq{tHooft_scale_inv}.)
The point of this kernel is that it satisfies
\beq
  \frac{iN_c}{\pi} \int_0^1\!\! dx\, G_0(q^2,x) \, (q^2 - \hat{T}\!*) 
     G_0(q^2, x) 
  = \langle P\,P \rangle (q)  \>.
\eql{GG=PP}
\eeq
What is the 3d `dual' of this kernel?  Let us 
use our transform \eq{thetransform} to find it.  First, let us take
the scale-invariant limit, in which $G_0$ becomes
\beq
  G_0(q^2,x) \too
    m_q \int_0^\infty\!\! \frac{dm^2}{2\pi^2} \, 
    \frac{\phi(m^2 x)}{q^2-m^2} \int_0^\infty\! \frac{dy}{y} \, \phi(m^2 y) \>,
\eeq
where the factor $2\pi^2$ comes from the fact that the modes 
$n=0,2,4,\cdots$ have spacing $2\pi^2\Lambda^2$.
Now, following our transform, let us define $\bar{G}_0(q^2,z)$ via
\beq 
  G_0(q^2,x) 
  = \int \frac{dz}{z} 
      \left[ \del_z \phi\!\left( \frac{\pi^2 z^2}{4x} \right) \right] 
      \bar{G}_0(q^2, z)  \>,
\eeq
and compute the left-hand side of \eq{GG=PP} in terms of $\bar{G}_0$.  
It has two pieces, the $q^2$ piece and the $\hat{T}$ piece.
First, the $q^2$ piece becomes
\beq
  && \int\! dx \, G_0(q^2,x) \, q^2 G_0(q^2, x)  \nn\\
  &=& \int\! \frac{dz}{z} \frac{dz'}{z'} \, 
               \bar{G}_0(q^2, z) \, q^2 \bar{G}_0(q^2, z') 
      \int\! dx \left[ \del_z \phi\! \left( \frac{\pi^2 z^2}{4x} \right) \right] 
                \left[ \del_{z'} \phi\! \left( \frac{\pi^2 z'^2}{4x} \right)
                \right]  \nn\\
  &=& \pi^2 \int\! dz dz' \, \bar{G}_0(q^2, z) \, q^2 \bar{G}_0(q^2, z')
      \int\! \frac{dx}{2x^2} [\sin(\pi z^2/4x) \, \sin(\pi z'^2/4x) 
                              + O(\widetilde{m}_q)]  \nn\\
  &=& \pi^2 \int\! \frac{dz}{2z} \,\bar{G}_0(q^2, z) \,q^2 \bar{G}_0(q^2, z)
      + O(\widetilde{m}_q)  \>,
\eql{Gq^2G}
\eeq
where we have used 
$\int_0^\infty\!dx\, \sin[ax] \sin[bx] = \frac{\pi}{2} \delta (a-b)$ in the last
step. 
On the other hand, since $\phi(\frac{\pi^2 z^2}{4x}) \to 0$ as $z \to 0$, 
$G_0(q^2,x)$ may also be written as,
\beq
  G_0(q^2,x) = - \int\! dz \,\phi\! \left( \frac{\pi^2 z^2}{4x} \right) \, 
                 \del_z\!\!\left( \frac{\bar{G}_0(q^2, z)}{z} \right),
\eeq
and so the $\hat{T}$ piece becomes
\beq
  && \int\! dx \, G_0(q^2,x) \> \hat{T}\!*\!G_0(q^2, x) \nn\\
  &=& \pi^2 \int\! \frac{dz \, dz'}{4} \, 
        \del_z\!\! \left( \frac{\bar{G}_0(q^2, z)}{z} \right) \, z^{\prime 2} \, 
        \del_{z'}\!\! \left( \frac{\bar{G}_0(q^2, z')}{z'} \right)
      \int\! \frac{dx}{x^2} \, \phi\! \left( \frac{\pi^2 z^2}{4x} \right) \, 
        \phi\! \left( \frac{\pi^2z'^2}{4x} \right) \nn\\
  &=& \pi^2 \int\! dz \, \frac{z}{2} \, 
        \del_z\! \left( \frac{\bar{G}_0(q^2, z)}{z} \right) \, 
        \del_{z}\! \left( \frac{\bar{G}_0(q^2, z)}{z} \right)  
      \eql{GTG}\\
  &=& \pi^2 \int\! \frac{dz}{2z} 
         \left( [\del_z \bar{G}_0(q^2, z)] \, [\del_{z} \bar{G}_0(q^2, z)] 
               -\frac{1}{z^2} \bar{G}_0(q^2, z) \, \bar{G}(q^2, z)
         \right)
      +\frac{\pi^2}{2\epsilon^2} \, 
         \bar{G}_0(q^2, \epsilon) \, \bar{G}_0(q^2, \epsilon)  \>.\nn  
\eeq
Thus, combining \eq{Gq^2G} and \eq{GTG}, we get 
\beq
&& \langle P\,P \rangle(q) 
   = -i \frac{\delta^2 S_{\rm AdS}}{\delta J_P(-q) ~\delta J_P(q)} 
   = \frac{iN_c}{\pi} \int\! dx \, G_0(q^2,x) \, (q^2 - \hat{T}\!*) G_0(q^2, x)
        \nn\\
&=& i \pi N_c \int\! \frac{dz}{2z} 
      \left( \bar{G}_0(q^2, z) \, q^2 \bar{G}_0(q^2, z)
            -[\del_z \bar{G}_0(q^2,z)]^2 
            +\frac{1}{z^2} [\bar{G}_0(q^2,z)]^2 
      \right)  \nn\\
&& +\frac{i \pi N_c}{2\epsilon^2} [ \bar{G}_0(q^2, \epsilon) ]^2  \>.
\eeq
This indeed implies that $\bar{G}_0$ is the bulk-to-boundary propagator for
the bulk field $X$ with the bulk action precisely equal to 
\eq{conformal-X-action}, with the additional boundary term 
$\sim \frac{1}{\epsilon^2} \cal{X}^\dagger \cal{X}$.   
The boundary term is just an indication that 
$\bar{G}_0(-Q^2, z) \sim z K_0(Qz)$ 
(i.e.~without being divided by $\epsilon K_0(Q\epsilon)$), 
which is just an alternative convention for the normalization of 
the field from that of \eq{conformal-X-solution}.  Therefore, we have found
that the transform \eq{thetransform} directly maps the bulk-to-boundary
propagator $\bar{G}$ to the Green's function $G$ of the 't Hooft equation!

\section{Towards Full Implementation of Conformal Symmetry Breaking}
\label{sec:towards}

Thus far we have discussed the 3d dual of the 't Hooft model near its 
conformal limit.  What can we expect the dual of the full confining theory 
to look like?  First, We have seen that 3d equations have essentially followed
from the 't Hooft equation.  On the other hand, the simplest basis of 3d fields
consists of fields dual to primary operators.  Thus, it is natural to express
the 't Hooft equation \eq{tHooftEq} in the basis of primary operators, which
is spanned by the Legendre Polynomials as we have seen in section 
\ref{sec:tHooft-correlators}.  In this basis, the 't Hooft operator $\hat T$ 
becomes
\beq
  \hat{T}_{kk'} 
    = (2k'-1) \int_0^1\!\! dx \int_0^1\!\! dy \, P_{k-1}(2x-1) 
        \left[ \frac{m_q^2 -   \Lambda^2}{x(1-x)}\delta(x-y) 
              -\hat{{\rm P}} \frac{\Lambda^2}{(x-y)^2}
        \right] P_{k'-1}(2y-1) \,.  \nn\\
\eeq
Then, the 't Hooft equation \eq{tHooftEq} becomes a matrix equation 
\beq
  \sum_{k'} \hat{T}_{kk'} \, M_{k'n} = m_n^2 \, M_{kn}  \>,
\eql{tHooftEq-matrix}
\eeq
where $M_{k,n}$ are the moments defined in \eq{moments}.  This is not the only
way to discretize the 't Hooft equation, but this is the most natural one
suggested by AdS/CFT.

To extract information about how bulk fields mix in the 3d action, 
we would like to have kernels of the 't Hooft equation which get mapped 
to `bulk-to-boundary' propagators.  We have seen this explicitly for 
$\langle PP \rangle$ in section \ref{sec:transform}.  So, generalizing the
kernel $G_0$ to all other primary operators, let us define  
\beq
  G_k(q^2,x) 
    \equiv \sum_{n}\!
             \frac{\phi_n(x)}{q^2-m_n^2} 
             \int_0^1\!\! dy \, P_{k-1}(2y-1) \,\phi_n(y)
         = \sum_{n}\! \frac{\phi_n(x)}{q^2-m_n^2} M_{k,n}  \>.
\eeq
Like $G_0$, this satisfies
\beq
  \frac{iN_c \, q_+^{2k}}{\pi} 
  \int_0^1\!\! dx\, G_k(q^2,x) \, (q^2 - \hat{T}\!*) G_k(q^2, x) 
  = \langle L_{k+} \, L_{k+} \rangle (q)  \>.
\eql{GG=LL}
\eeq
In the basis of primary operators, the 't Hooft equation implies
\beq
  \sum_{k'} \left( \sqrt{\frac{2k'-1}{2k-1}} q^2 \delta_{kk'} - \hat{T}_{kk'} 
            \right) 
            G_{k'}(q^2,x) 
 = P_{k-1}(2x-1)  \>.
\eeq
Therefore, the matrix $\hat{T}_{kk'}$ can be thought of as containing the
information regarding the mixing of 3d fields dual to primary operators,
following conformal symmetry breaking.  (Note that the 't Hooft operator 
$\hat{T}_{kk'}$ is proportional to $\Lambda^2$ in the $m_q \to 0$
limit.)

The hope is then that one could transform the above equations in $x$ into
a set of coupled 3d equations of motion.  Indeed, one can write down an
abstract formula for the transform for the full theory
\beq
\label{exacttrans}
F(x,z) = \sum_n \phi_n(x) \tilde{\phi}_n(z),
\eeq
where $\tilde{\phi}_n(z)$ are the bulk KK-modes (i.e.~the normalizable 
solutions to the set of coupled 3d equations). The resulting 3d equations 
of motion would encode all information about conformal symmetry breaking,
including all possible mixings of bulk fields.
In addition, they should tell us how the Regge-like spectrum 
$m_n^2 \propto n$ could arise as a consequence of the mixings, 
and ultimately at least some qualitative features of the backgrounds 
causing all the mixings.  Some hint of the effective result of the mixing 
can already be seen from an approximate form of eq.(\ref{exacttrans}) valid 
for large $n$
\beq
  F(x,z) \sim \sum_n \sqrt{2} \, \cos[\pi n \Lambda^2 x] \,
              z L_n \!\!\left( \frac{\pi^2 \Lambda^2 z^2}{4} \right)  ,
\eeq
where $L_n$ are the Laguerre polynomials.  This form follows
from the fact that the Laguerre polynomials provide the right spectrum at
large $n$, and under the previously used conformal limit, 
$\Lambda \rightarrow 0$ and $n \rightarrow \infty$ with
$m^2 \sim \pi^2 \Lambda^2 n$ fixed, 
$L_n(\frac{m^2 z^2}{4n}) \rightarrow J_0(mz)$.
Transforming the 't Hooft equation for a meson of sufficiently large $n$, 
by this $F(x,z)$, will therefore yield the equation of motion resulting from 
a background similar to \cite{linear-confinement}.  In other words, the 
`dilaton' profile seems to appear as an effective background which 
approximates the effect of field mixing for the highly excited modes.

\section{Conclusion}
\label{sec:conc}

In this paper we have taken some steps towards describing the 3d dual to 2d
QCD at large $N_c$. In the conformal limit we have proposed the form of the
quadratic 3d action for the duals of primary operators.  We have also included
the leading effects of conformal symmetry breaking. 
We also proposed a transform (in the conformal limit) which relates
the 't Hooft wavefunctions to the bulk modes, therefore enabling us to
map the 't Hooft equation to the equation of motion for a bulk
scalar.
Some conjectured features 
of the full dual and the transform at the quadratic level were provided, 
and we hope to report on the particulars in a future paper.  

There are several intriguing open questions.  Though we have only described the
quadratic part of the action, one may use the transform to derive the cubic 
terms as well at leading order in $N_c$ (at least in the conformal limit). 
Indeed, at large $N_c$, on the 2d side there are expressions for the three-point
correlators in terms of the parton wavefunctions \cite{Gross}.  These may be
transformed into bulk cubic vertices in the AdS region of the background.  It
would be interesting to see how these compare to known actions from
supersymmetric duals.  One could also study deep inelastic scattering 
at leading order in $N_c$ and compare with \cite{DISads}.  Finally, it 
would be interesting to study the effect of quark masses on the 3d dual, 
perhaps also taking the heavy quark limit.

\section*{Acknowledgment}

We thank Mithat \"Unsal for his collaboration at the initial stage of 
the project.  T.O.~thanks Yuko Hori for pointing out some crucial errors in 
the lengthy calculations in section \ref{sec:dim-2} and appendix 
\ref{app:spin2-const-eq}, and for useful discussions about the materials
therein.  We thank Rich Brower for introducing us to the equation 
\eq{tHooft_scale_inv}.  We also thank Andreas Karch, Matt Reece, and Raman Sundrum for 
their comments on the draft.  E.K.~was supported in part by the Department of Energy grant 
no. DE-FG02-01ER-40676, by the NSF CAREER grant PHY-0645456, and by the Alfred P. Sloan Fellowship.
T.O.~has been supported by NSF grant NSF-PHY-0401513, 
by the Johns Hopkins Theoretical Interdisciplinary Physics and Astrophysics Center,
and by the Maryland Center for Fundamental Physics.

\section*{Appendices} 
\renewcommand{\thesection}{\Alph{section}}
\setcounter{section}{0}
\setcounter{subsection}{0}
\setcounter{subsubsection}{0}
%

\section{The Primary Operators}
\label{app:primary-op}

In this appendix, we will compile a list of all primary single trace operators in the 't
Hooft model, except for those which vanish by the equations of motion 
in the conformal limit.

By definition primary operators are operators that transform covariantly under
conformal transformations, just like tensor operators are ones that transform
covariantly under Lorentz transformations.  Since the Lorentz group is a
subgroup of the conformal group, all primary operators are Lorentz tensors 
(but the converse is not true).  In $1+1$ dimensions tensor components can 
be handled most efficiently in terms of the light-cone coordinates 
$x^\pm = (x^0 \pm x^1)/\sqrt{2}$, where the metric is simply 
$ds^2 = 2 dx^+ dx^-$.  Aside from {\it parity} $x^+ \leftrightarrow x^-$ 
and {\it time-reversal} $x^+ \leftrightarrow - x^-$, a Lorentz transformation
is given by
$x^\pm \too e^{\pm\lambda} x^\pm$ with a real parameter $\lambda$ (i.e.~the 
`rapidity').  Then, $\del_\pm = (\del_0 \mp \del_1)/\sqrt{2}$ transform as
$\del_\pm \to e^{\mp\lambda} \del_\pm$, while left-moving and right-moving 
spinors $\psi_+$ and $\psi_-$ transform as a `square-root' of $\del_+$ and 
$\del_-$, namely, as $\psi_\pm \to e^{\mp \lambda/2} \psi_\pm$.  Note that the
standard kinetic terms for $\psi_+$ and $\psi_-$,
\beq
  \int\! dx^+ dx^-\, \sqrt{2}\, \psi_+^\dagger \del_- \psi_+
  \quad, \quad
  \int\! dx^+ dx^-\, \sqrt{2}\, \psi_-^\dagger \del_+ \psi_-  \>,
\eql{fermion-kin}
\eeq
are manifestly invariant under these transformations.

However, \eq{fermion-kin} are clearly invariant under more general
transformations, or {\it conformal transformations,}
\beq
  x^+ \too x^{\prime+} = f^+(x^+)
  \quad, \quad
  x^- \too x^{\prime-} = f^-(x^-)  \>,
\eeq
where $f^\pm$ are two independent, arbitrary functions, provided that we also 
let $\psi_\pm$ transform as
\beq
  \psi_\pm (x) \too \psi'_\pm (x') 
  \equiv \left| \frac{df^\pm}{dx^\pm} \right|^{-1/2} \psi_\pm (x)  \>.
\eeq
This symmetry group is enormous, much larger than the isometry group
of AdS$_3$, which only has six generators.  For the purpose of AdS$_3$/CFT$_2$,
therefore, we are only interested in {\it special conformal transformations}, 
a subset of the above transformations, with globally defined generators.  For 
infinitesimal transformations, this means we should restrict $f^\pm$ to 
just {\it quadratic} functions,
\beq
  f^+(x^+) &=& x^+ + \alpha^+ + (\delta+\lambda) x^+ + \epsilon_+ (x^+)^2  
               \>,\nn\\
  f^-(x^-) &=& x^- + \alpha^- + (\delta-\lambda) x^- + \epsilon_- (x^-)^2  \>,  
\eql{f-forms}
\eeq
which depend on six (infinitesimal) parameters $\alpha^\pm$, 
$\lambda$, $\delta$, and $\epsilon_\pm$.  Clearly, $\alpha^\pm$ and $\lambda$
just parameterize Poincar\'e transformations.  Among the three `new' parameters, 
$\delta$ induces a {\it dilation} $x^\pm \too (1+\delta)x^\pm$, while 
$\epsilon_\pm$ induce a {\it conformal boost} 
$x^\pm \too (1 + \epsilon_\pm x^\pm) x^\pm$. 

Now we are ready to write down the general transformation law for any
primary operators.  First, let us define our notation.  Say, we have an
operator with $n_+$ lower $+$ indices and $n_-$ lower $-$ indices, counting
spinorial $\pm$ as half.  (For example, $(n_+, n_-) = (1,0)$ for $\del_+$, 
while $(n_+, n_-) = (3/2,1/2)$ for $\psi_- \del_+\psi_+$.)  We then define the 
{\it spin} $s$ of the operator by $s = n_+ - n_-$.  Our convention for 
scaling dimensions is such that a $\del$ has scaling dimension one under the
dilation.  Now, 
if an operator ${\cal O}_{\Delta, s}$ with scaling dimension $\Delta$ and 
spin $s$ is a primary operator, then it should transform in the same way as 
$(\psi_+)^{\Delta +s} (\psi_-)^{\Delta -s}$.  Namely,
\beq
  {\cal O}_{\Delta,s}(x) 
  \too {\cal O}'_{\Delta,s} (x')
       \equiv \left| \frac{df^+}{dx^+} \right|^{-\frac{\Delta +s}{2}}
              \left| \frac{df^-}{dx^-} \right|^{-\frac{\Delta -s}{2}}
              {\cal O}_{\Delta,s} (x)  \>,
\eql{prime-op-trans}
\eeq
where $f^\pm$ have the form \eq{f-forms}.  Hereafter we will refer to this
simply as `conformal transformation' without `special'.

\subsection{$U(1)_A$-Neutral Primary Operators}

These operators are further divided into two classes, the $L$ type and the $R$
type.  The analysis of the $R$ type goes exactly parallel to that of the $L$
type, so here we will just discuss the operators of the $L$ type,
which are a linear combination of the operators of the 
form $({\cal D}_1 \psi_+^\dagger) ({\cal D}_2 \psi_+)$
where ${\cal D}_{1,2}$ are some powers of $\del_+$.  It cannot contain $\del_-$,
since it would then vanish by the equation of motion $\del_- \psi_+ = 0$ in the
conformal limit.

Obviously, the lowest-dimensional $L$ type operator is 
$\psi_+^\dagger \psi_+$, which has $\Delta = 1$ and $s=1$.  This is of course
the $+$ component of the $U(1)_L$ Noether current.  
The next lowest one must be a linear combination of 
$\psi_+^\dagger \del_+ \psi_+$ and $(\del_+\psi_+^\dagger) \psi_+$.
Under a conformal transformation, they transform as
\beq
  \psi_+^\dagger \del_+ \psi_+
    &\too& J_+^2 \, \psi_+^\dagger \del_+ \psi_+
           + J_+^{3/2} (\del_+ J_+^{1/2}) \, \psi_+^\dagger \psi_+
      \>,\nn\\
  (\del_+\psi_+^\dagger) \psi_+
    &\too& J_+^2 \, (\del_+\psi_+^\dagger) \psi_+
           + J_+^{3/2} (\del_+ J_+^{1/2}) \, \psi_+^\dagger \psi_+  \>,
\eeq
where $J_+ \equiv | df^+/dx^+ |^{-1}$.  Note that if we subtract one of these
from the other, it agrees with the form \eq{prime-op-trans}.  
So the primary operator must be the following linear combination:
\beq
  \psi_+^\dagger \del_+ \psi_+ - (\del_+\psi_+^\dagger) \psi_+  \>,
\eeq
which has $\Delta=2$ and $s=2$.  This is nothing but the $++$ component of
the energy-momentum tensor.  (For $\Delta=2$ and $s=0$, the 
combination $\psi_+^\dagger \del_- \psi_+ - (\del_-\psi_+^\dagger) \psi_+$ 
does transform as a primary operator, but, as we mentioned already, 
this vanishes by the equation of motion in the conformal limit.) 

Proceeding to the next level, we have to find an appropriate linear 
combination of
$\psi_+^\dagger \del_+^2 \psi_+$, $(\del_+ \psi_+^\dagger) \del_+ \psi_+$, 
and $(\del_+^2 \psi_+^\dagger) \psi_+$.  Repeating the above exercise, 
we find that again there is a unique combination which obeys the law 
\eq{prime-op-trans}:
\beq
  \psi_+^\dagger \del_+^2 \psi_+ - 4(\del_+ \psi_+^\dagger) \del_+ \psi_+
   + (\del_+^2 \psi_+^\dagger) \psi_+  \>,
\eeq
which has $\Delta=s=3$.  At the next level, one finds that the coefficients of
$\psi_+^\dagger \del_+^3 \psi_+$, $(\del_+\psi_+^\dagger) \del_+^2 \psi_+$, 
$(\del_+^2 \psi_+^\dagger) \del_+ \psi_+$, 
$(\del_+^3 \psi_+^\dagger) \psi_+$ are $1$, $-9$, $9$, $-1$, respectively.
Thus, the coefficients are given by the square of the binomial
coefficients with alternating signs.  Therefore, the $L$-type primary
operator with $\Delta=s=k$ is given by
\beq
  L_{k+} 
    &\equiv& i^{k-1} \sqrt{2} 
             \sum_{m=0}^{k-1} ({}_{k-1} {\rm C}_m)^2 \, (-1)^m \,
               (\del_+^m \psi_+^\dagger) \, \del_+^{k-1-m} \psi_+  \nn\\
         &=& \sum_{m=0}^{k-1} ({}_{k-1} {\rm C}_m)^2 \,
               [(-i\del_+)^m \overline{\psi}] \gamma_+ \, 
               (i\del_+)^{k-1-m} \psi  \>,
\eeq
where ${}_n {\rm C}_m \equiv n!/[m! \, (n-m)!]$, and 
$\gamma_+ = (\gamma_0 + \gamma_1)/\sqrt{2}$.

\subsection{$U(1)_A$-Charged Primary Operators}

Clearly, the lowest-dimensional primary operators in this class are
$\psi_+^\dagger \psi_-$ and its Hermitian conjugate.  With one $\del_+$,
the only combination that does not vanish by the equations of motion is
$(\del_+\psi_+^\dagger) \psi_-$ (and its Hermitian conjugate).  However, this
does not transform as \eq{prime-op-trans} because it gives an extra term
containing 
$\del_+ J_+$.  Since this is the only operator with $\Delta=2$ and $s=1$ that
does not vanish by the equations of motion, there is no way to cancel this
extra term.  (Actually, even if we forget about the equations of motion,
$\psi_+^\dagger \del_+\psi_-$ still would not help us since it would only give
$\del_+ J_-$ instead of $\del_+ J_+$.)  This problem persists for 
$(\del_+^p \psi_+^\dagger) \del_-^q \psi_-$ with any $p$, $q$.  Thus, we
conclude that $\psi_+^\dagger \psi_-$ and its Hermitian conjugate are the only 
(non-vanishing) primary operators in this class.

\section{2D Calculation of 2-Point Correlators}
\label{app:correlator-calc}

In this appendix, we derive the formulae \eq{full-SS}, \eq{full-PP}, and
\eq{full-O_k-O_l}.  We essentially follow the method in \cite{Gross} and
generalize it to include all the primary operators.  
Throughout this appendix, we choose the units where $\Lambda = 1$. 

\subsection{The Feynman Rules}

The Feynman rules in the 't Hooft double-line notation are:
\begin{itemize}
\item{
 The gluon propagator:
 \beq
   \lower0.25em\hbox{\includegraphics[width=1.2in]{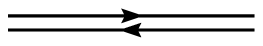}}
   \>\> = \>
   \frac{\pi}{N_c} 
   \biggl( \delta^a_c \delta^d_b - \frac{1}{N_c} \delta^a_b \delta^d_c \biggr)
   \frac{i}{k_-^2} \quad , \quad \abs{k_-} > \lambda  \>,
 \eql{gluon-prop}
 \eeq
 where $\lambda$ is an IR cutoff, and $a,b,\cdots$ label color.  
 The second term in the bracket is subleading
 in $1/N_c$ expansion, and hence not used in this paper.
}
\item{
 The quark propagator:
 \beq
   \hbox{\includegraphics[width=1.2in]{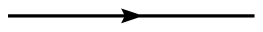}}
   \>\> = \>
   \frac{i(\gamma_+ p_- + \gamma_- p_+ + m_q)}{2p_+ p_- - m_q^2 +i\varep}  \>.
 \eeq
}
\item{
 The quark-quark-gluon vertex:
 \beq
   \hbox{\includegraphics[width=1in]{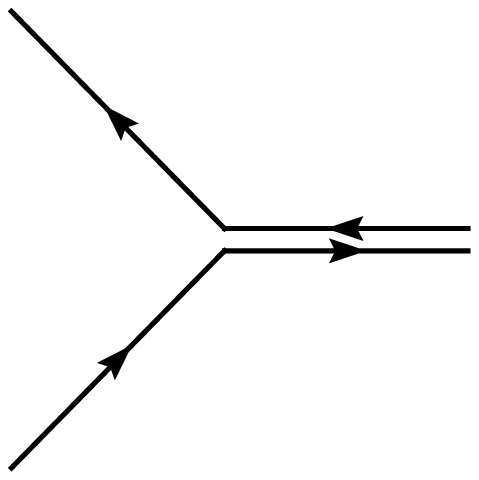}} 
   \raise3em\hbox{$\quad = \> -i \gamma_- \>.$}
 \eeq
}
\end{itemize}
We will choose the light-cone gauge $A_-=0$ in the following calculations.
The advantage of this gauge is that all gluon self-couplings vanish identically.

\subsection{The Quark Self-Energy}

At the leading order in the $1/N_c$ expansion, only the quark propagator gets
quantum corrections; the gluon propagator and the quark-quark-gluon vertices
remain unchanged.

Since the 1PI quark self-energy is proportional to $\gamma_-$ in the $A_-=0$
gauge, we define
\beq
  \text{(The 1PI quark self-energy)} \equiv -i \Sigma(p) \gamma_-  \>.
\eeq
Then, the exact full quark propagator can be written as
\beq
   \frac{i[ p_- \gamma_+ + \left( p_+ - \Sigma(p) \right) \gamma_- + m_q ]}
        {2 p_- \left( p_+ - \Sigma(p) \right) - m_q^2 + i\varep}  \>.
\eql{full-quark-prop}
\eeq
Now, at the leading order in $1/N_c$, only the ``rainbow'' diagrams contribute
(see figure \ref{fig:Sigma}).
\FIGURE[t]{\epsfig{file=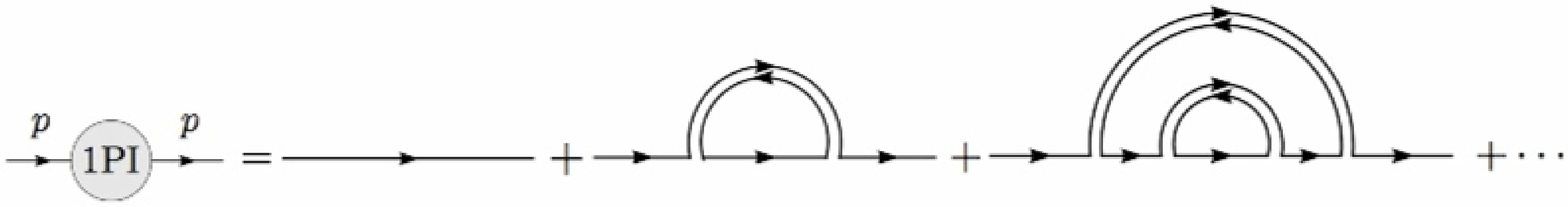, width=5.5in}\caption{The quark self-energy at the leading in $1/N_c$.}\label{fig:Sigma}}

Also, by inspecting the diagrams, we see that $\Sigma(p)$ only depends on $p_-$.
Therefore, we have
\beq
  -i\Sigma(p_-) = \frac{1}{4\pi} \int\! dk_+ dk_- \,
                    \frac{1}{\left[ (k_- - p_-)^2 \right]_\lambda} \,
                    \frac{1}{k_+ - \Sigma(k_-) - \frac{m_q^2}{2k_-}
                             + i\varep\, {\rm sgn}(k_-)}
\eeq
where ${\rm sgn}(k_-) \equiv k_-/|k_-|$ and the notation $[\cdots]_\lambda$ is
meant to remind us of the IR cutoff on the gluon propagator \eq{gluon-prop}.  
The $k_+$ integral here is log divergent.  
We choose to remove the divergence by imposing a symmetric cutoff on $k_+$
(i.e.~$\abs{k_+} \leq \Lambda$) {\it after} shifting $k_+$ as 
$k_+ \too k_+ + \Sigma(k_-) + m_q^2/2k_-$ to eliminate the terms 
$-\Sigma(k_-) - m_q^2/2k_-$ in the denominator.
Having done so, we get
\beq
  \Sigma(p_-) = \frac{{\rm sgn}(p_-)}{2\lambda} - \frac{1}{2p_-}  \>.
\eeq
%

\subsection{The Quark-Antiquark `Scattering' Matrix}

Consider the diagrams in figure \ref{fig:T}.
Here, we are {\it not} trying to calculate a scattering amplitude (quarks can
never be put on-shell anyway)---rather, since such diagrams will often appear as 
part of larger diagrams, it is convenient to evaluate them once and for all.

At the leading order in $1/N_c$, the only way for the quark and antiquark to exchange gluons is in the ``ladder'' fashion where all gluons just go vertically
connecting the quark and antiquark, and no two gluons ever cross.  
All diagrams of this type have a $\gamma_-$ for each quark line, and one color flows in along the upper-left line and flows out along the lower-left line, and another color, independent of the first one, flows in along the lower-right line and flows out along the upper-right line.

\FIGURE[t]{\epsfig{file=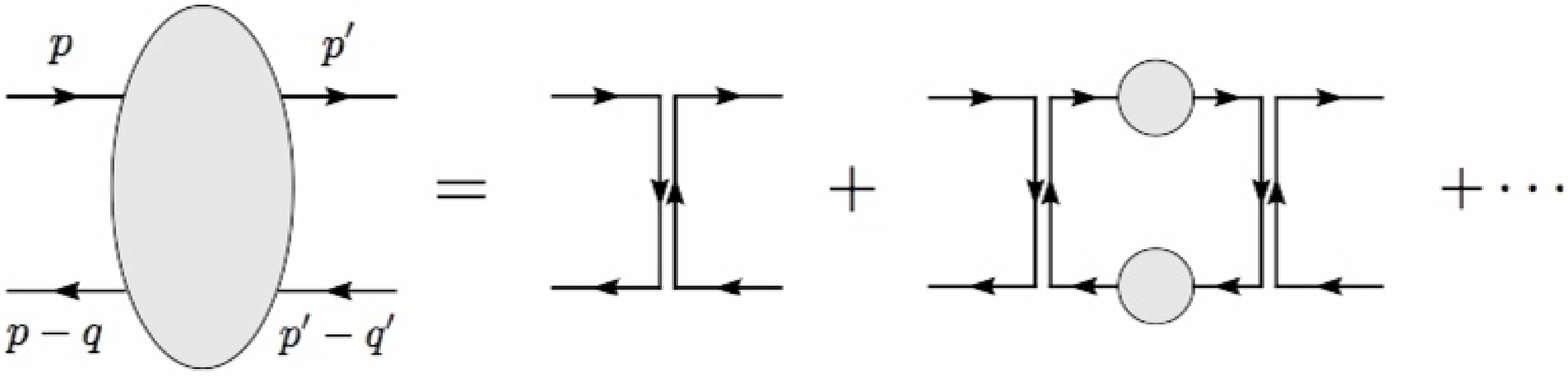, width=5.5in}\caption{The quark-antiquark `scattering' at the leading order in $1/N_c$. A gray circle represents the full quark propagator \eq{full-quark-prop}.}\label{fig:T}}

Let $T(p, p'; q)$ be the sum of all such ladder diagrams.  (The color indices, the flavor indices, and the factor of $\gamma_- \otimes \gamma_-$ are
suppressed.)
Then, we have
\beq
  T(p,p';q)
    = -\frac{i\pi}{N_c} \frac{1}{\left[ (p_- - p'_- )^2 \right]_\lambda}
      + \frac{i}{4\pi}
          \int\! dk_- \, \frac{1}{\left[ (p_- - k_- )^2 \right]_\lambda}
            \Phi(k_-,p';q)  \>,
\eql{Tbootsstrap}
\eeq
where
\beq
\eql{Phi-def}
  \Phi(p_-, p'; q) \equiv \int\! dp_+ \, S(p) \, S(p-q) \, T(p,p';q)  \>,
\eeq
and
\beq
  S(p) \equiv \frac{1}{p_+ - \Sigma(p_-) - \frac{m_q^2}{2p_-}
                       + i\varep\, {\rm sgn}(p_-)}  \>.
\eeq
Since by definition $\Phi (p_-, p'; q)$ does not depend on $p_+$, 
\eq{Tbootsstrap} tells us that $T(p, p'; q)$ does not depend on $p_+$ either.  
So, the $p_+$ integral in \eq{Phi-def} converges.  
For $0< p_- / q_- < 1$ we get
\beq
\eql{Phi-T}
  \Phi(p_-, p'; q)
    = \frac{2\pi i \, {\rm sgn}(q_-) \, T(p_-, p'; q)}
           {q_+ + \Sigma(p_- - q_-) - \Sigma(p_-)
            + \frac{m_q^2}{2(p_- - q_-)} - \frac{m_q^2}{2p_-}
            + i\varep \,{\rm sgn}(q_-) }  \>,
\eeq
while for $p_- / q_- \geq 1$ or $p_- / q_- \leq 0$ we get
\beq
  \Phi(p_-, p'; q) = 0  \>.
\eeq
Then, for $0< p_- / q_- < 1$, putting \eq{Tbootsstrap}
into \eq{Phi-T} gives
\beq
  &&  \frac{m_q^2-1}{\hat{p} (1 - \hat{p})} \, \Phi(p_-, p'; q)
     -\hat{{\rm P}} \!\!\int_0^1
        \frac{\Phi(q_- x, p'; q)}{(x - \hat{p})^2} \,dx  \nn\\
  &=& -\frac{4\pi^2}{N_c \abs{q_-}}
       \frac{1}{\left[ (\hat{p} - \hat{p}')^2 \right]_{\hat{\lambda}}}
      + (q^2 + i\varep) \,\Phi(p_-, p'; q)  \>,
\eeq
where $\hat{p} \equiv p_-/q_-$, $\hat{p}' \equiv p'_-/q_-$, and
$\hat{\lambda} \equiv \lambda / \abs{q_-}$.
For $0<x<1$, this can be solved in terms of the 't Hooft wavefunction 
$\phi_n (x)$ satisfying the 't Hooft equation \eq{tHooftEq}.
For $x \leq 0$ or $\geq 0$, we set $\phi_n(x)=0$ by definition.
Then, we have
\beq
  \Phi(p_-, p_-'; q)
  = \frac{4\pi^2}{N_c \abs{q_-}}
    \sum_n \frac{1}{q^2 - m_n^2 + i\varep} \, \phi_n(\hat{p})
      \int_0^1\!\! dx \, 
        \frac{\phi_n^* (x)}{\left[ (x - \hat{p}')^2 \right]_{\hat{\lambda}}}
\>,
\eeq
for all real values of $p_-$.

Therefore, we finally get
\beq
   T(p_-, p'_-; q)
  = -\frac{i\pi}{N_c} \frac{1}{\left[ (p_- - p'_- )^2 \right]_\lambda}
    +\frac{4\pi}{N_c} 
     \sum_n \frac{i}{q^2 - m_n^2 + i\varep}
       \frac{\psi_n (\hat{p}, q_-) \, \psi_n^* (\hat{p}', q_-)}{\lambda^2}
\eeq
where
\beq
  \psi_n (x, q_-) \equiv
    \frac{\lambda}{2|q_-|} \int_0^1\!\! dy \, 
    \frac{\phi_n (y)}{[(y-x)^2]_{\hat{\lambda}}}  \>.
\eql{psi_n}
\eeq
For $0<x<1$, the 't Hooft equation \eq{tHooftEq} tells us that $\psi_n$ is 
equal to $\phi_n$ up to an $O(\lambda)$ correction:    
\beq
  \psi_n (x, q_-)
    =  \left[ 1 - \frac{\lambda}{2\abs{q_-}}
                  \left( m_n^2 - \frac{m_q^2-1}{x(1-x)} \right)
                  \right] \phi_n (x)
    = \phi_n(x) + O(\lambda)  \>.
\eeq
For $x<0$ or $>1$, we can remove the IR cutoff in the integrand in \eq{psi_n},
so $\psi_n$ can be written as
\beq
  \psi_n (x, q_-) 
  = \frac{\lambda}{2|q_-|} \int_0^1\!\! dy \, \frac{\phi_n (y)}{(y-x)^2}  
  = O(\lambda)  \>.
\eeq
%

\subsection{Computation of 2-point Correlators}

In our gauge $\langle R_{n-} R_{n-} \rangle$ is the easiest one to compute.  
First, let us define
\beq
  R_{k,\ell} \equiv [ (-i\del_-)^k \overline{\psi} ] 
                    \gamma_- (i\del_-)^\ell \psi \>,
\eeq
so that
\beq
  R_{n-} = \sum_{k=0}^{n-1} ( {}_{n-1} C_k )^2 \, R_{n-1-k,k}  \>.
\eeq
Then, in terms of $T(p_-, p'_- ;q)$ calculated above, the correlator can be
expressed as in figure \ref{fig:2pt}.
The simple quark-loop diagram without a $T$ blob will vanish in the limit of 
$\lambda \to 0$.  The contribution from the one with a $T$ blob is  
\beq
  &&  \langle R_{j,k} \, R_{\ell,m} \rangle (q)  \nn\\
  &=& -N_c^2 \int\! \frac{d^2p}{(2\pi)^2} \int\! \frac{d^2p'}{(2\pi)^2} \,
       S(p-q) \, S(P) \, S(p'-q) \, S(p')  \nn\\
  &&     \quad \times\,
       p_-^j \, (p_- - q_-)^k \,  p_-^{\prime m} \, (p'_- - q_-)^\ell \,
       T(p_-, p'_- ;q) \nn \,.   
\eeq
\FIGURE[t]{\epsfig{file=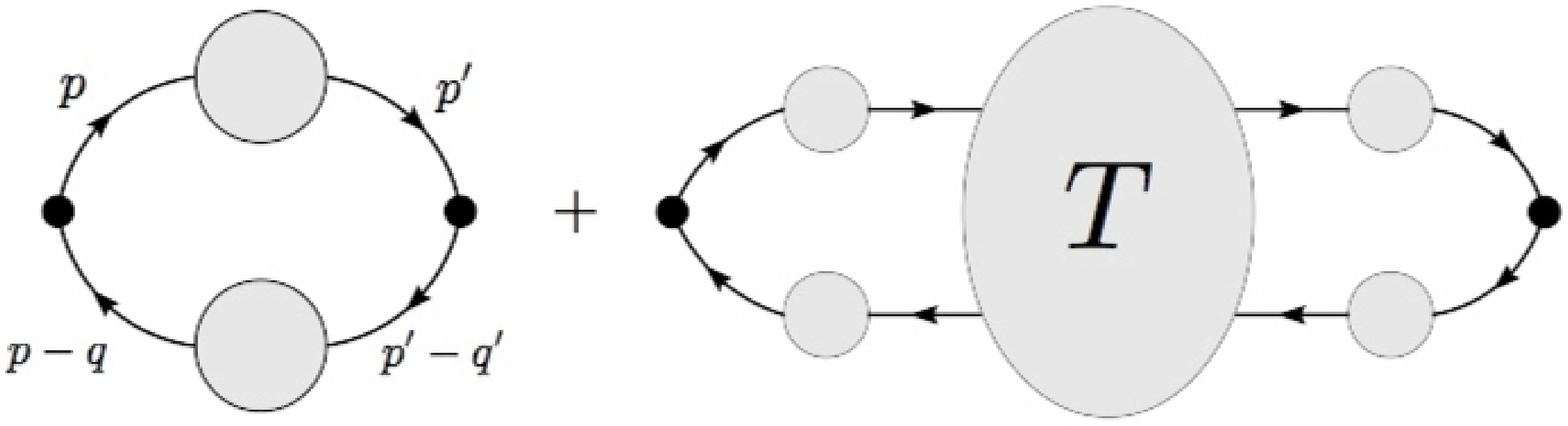, width=5.5in}\caption{Two-point correlators at the leading order in $1/N_c$. The big dots represent the operators $R_{k-}$, $R_{\ell -}$.}\label{fig:2pt}}
Performing $p_+$ and $p'_+$ integrals and taking the $\lambda \to 0$ limit,
we obtain
\beq
 && \langle R_{j,k} \, R_{\ell,m} \rangle (q)  \nn\\
 &=& \frac{N_c}{\pi} 
     \sum_n \frac{i q_-^{j+k+\ell+m+2}}{q^2 - m_n^2 + i\varep} 
     \left[ \int_0^1\!\! dx\, x^j (x-1)^k \, \phi_n(x) \right]
     \left[ \int_0^1\!\! dy\, y^m (y-1)^\ell \, \phi_n(y) \right]  \>.
\eeq
Now, notice that
\beq
  \sum_{k=0}^n ({}_n C_k)^2 x^{n-k} (x-1)^k
  = P_n(2x-1)  \>,
\eeq
where $P_n$ is the Legendre polynomial.  Therefore, we obtain 
\beq
  &&  \langle R_{k-} \, R_{\ell-} \rangle (q)  \nn\\
  &=& \frac{N_c}{\pi} 
      \sum_n \frac{i q_-^{k+\ell}}{q^2 - m_n^2 + i\varep} 
      \left[ \int_0^1\!\! dx\, P_{k-1}(2x-1) \, \phi_n(x) \right] \!\!
      \left[ \int_0^1\!\! dy\, P_{\ell-1}(2y-1) \, \phi_n(y) \right]
\eql{full-R_k-R_l}
\eeq
Translating this result to the $LL$ case is trivial.  Repeating the above steps
for $S = \overline{\psi} \psi$ and $P = \overline{\psi}i\gamma_3\psi$ 
to obtain \eq{full-SS} and \eq{full-PP} is also straightforward.

\section{Details of the Spin-2 Calculation}
\label{app:spin2-const-eq}

Due to the special role of the $z$ coordinate in the AdS$_3$ metric 
\eq{metric} and our choice of gauge \eq{grav-gauge}, it is necessary to treat 
`$3$' or `$z$' indices separately from `$\mu$' indices.  For this purpose, 
we need to know an explicit expression of the Christoffel symbol for the 
AdS$_3$ background $\hat{g}_{AB} = z^{-2} \, \eta_{AB}$:
\beq
  \hat{\Gamma}^A_{~BC}
  = -\frac{1}{z} ( \delta^3_B \delta^A_C + \delta^3_C \delta^A_B
                  - \hat{g}^{3A} \hat{g}_{BC} )  \>.
\eeq
(Note that $\hat{g}^{3A} \hat{g}_{BC}$ is actually independent of $z$, so the
whole $\hat{\Gamma}^A_{~BC}$ goes as $1/z$.)  Using this, we get the following
`rules' for $\nabla_{\!\!A} h_B^C$:
\beq
  \nabla_{\! 3}  h_B^C 
    &=& \del_3 h_B^C  \nn\\
  \nabla_{\!\alpha} h_\beta^\gamma 
    &=& \del_\alpha h_\beta^\gamma 
        - \frac{1}{z} \bigl( \delta_\alpha^\gamma h_\beta^3 
                            + \hat{g}_{\alpha\beta} h^{3\gamma} \bigr)  \nn\\
  \nabla_{\!\alpha} h_\beta^3 
    &=& \del_\alpha h_\beta^3 
        - z \bigl( h_{\alpha\beta} - \hat{g}_{\alpha\beta} h_3^3 \bigr)  \nn\\
  \nabla_{\!\alpha} h_3^3 
    &=& \del_\alpha h_3^3 
        + \frac{2}{z} h_\alpha^3  
\eql{rules}
\eeq
To derive \eq{S_EH} from \eq{L_EH-full}, we just use these formulae with the
gauge condition $h_{3M}=0$.  Next,  
in \eq{const_eq_1}-\eq{const_eq_3}, all the terms that are not multiplied by $c$
arise from varying $S_{\rm EH}$ with respect to $h_{3M}$.  It is a little more
work to get them because we must keep all terms linear in $h_{3M}$ until the 
end of the calculation.  But still it is not so laborious because $S_{\rm EH}$
itself is simple enough.

However, it is much more tedious to derive \eq{S_CS} and especially the 
$c$-dependent terms in \eq{const_eq_1}-\eq{const_eq_3}, because $S_{\rm CS}$
contains many more terms with more indices, so just classifying each index 
into `$\mu$' and `$3$' will give us a large number of terms.  Although this is
just a matter of algebra, we would like to mention a few things
that may help the reader verify those equations.

First, note that, for $g_{AB} = \hat{g}_{AB} + h_{AB}$, we have 
$\Gamma^A_{~BC} = \hat{\Gamma}^A_{~BC} + \delta\Gamma^A_{~BC}$ where
$\delta\Gamma^A_{~BC}$ consists of $h_{AB}$.
Then, correspondingly, we have 
$\Omega_{\rm CS} = \hat{\Omega}_{\rm CS} + \delta\Omega_{\rm CS}$ where 
we further split $\delta\Omega_{\rm CS}$ into two parts:
\beq
  \delta\Omega_{\rm CS}
  = \Omega_{\rm CS}^{(1)} + \Omega_{\rm CS}^{(2)}  \>,
\eeq
where, in terms of the matrix notation introduced in section \ref{sec:dim-2},
\beq
  \Omega_{\rm CS}^{(1)}
  \equiv \epsilon^{ABC} \,
         {\rm Tr}\bigl[\delta{\bf \Gamma}_{\!A} \,\del_B \hat{{\bf \Gamma}}_C 
                       +\hat{{\bf \Gamma}}_{\!A} \,\del_B \delta{\bf \Gamma}_C
                       +2\hat{{\bf \Gamma}}_{\!A} \,\hat{{\bf \Gamma}}_B \,
                         \delta{\bf \Gamma}_C \bigr]  \>,\nn\\
\eeq
and
\beq
  \Omega_{\rm CS}^{(2)}
  \equiv \epsilon^{ABC} \,
         {\rm Tr}\bigl[\delta{\bf \Gamma}_{\!A} \,\del_B \delta{\bf \Gamma}_C 
                       +2\hat{{\bf \Gamma}}_{\!A} \,\delta{\bf \Gamma}_B \,
                         \delta{\bf \Gamma}_C \bigr]  \>.
\eql{Omega-2}
\eeq
In the AdS$_3$ background, $\Omega^{(1)}_{\rm CS}$ is purely a total 
derivative.  This can be seen by first putting it in the following form:
\beq
  \delta\Omega_{\rm CS}^{(1)}
  = -\epsilon^{ABC} \, 
     \del_A {\rm Tr} \bigl[ \hat{{\bf \Gamma}}_B \,\delta{\bf \Gamma}_C \bigr]
    +\epsilon^{ABC} \, {\rm Tr} 
     \bigl[ \delta{{\bf \Gamma}}_{\!A} \,\hat{{\bf R}}_{BC} \bigr]  \>.  
\eeq
Then, note that since AdS$_3$ is maximally symmetric, we have 
$\hat{R}^A_{~BCD} = \hat{R} \, (\delta^A_C \, \hat{g}_{BD} 
- \delta^A_D \, \hat{g}_{BC})/6$ where $\hat{R}$ is the scalar curvature, 
which in turn implies that the second term in the above equation vanishes
identically.  Therefore, in the AdS$_3$ background, we have  
\beq
  \Omega_{\rm CS}^{(1)}
  = -\epsilon^{ABC} \, 
     \del_{A} {\rm Tr} \bigl[ \hat{{\bf \Gamma}}_B \,\delta{\bf \Gamma}_C \bigr]
       \>.
\eeq

So it comes down to evaluating $\Omega_{\rm CS}^{(2)}$.  Since it is already
quadratic in $\delta\Gamma^A_{~BC}$, we just need to express 
$\delta\Gamma^A_{~BC}$ to first order in $h_{AB}$:
\beq
  \delta\Gamma^A_{~BC}
  = \frac{1}{2} (\nabla_{\!B} h^A_C + \nabla_{\!C} h^A_B - \nabla^A h_{BC})
    + O(h^2)
\eql{deltaGamma}
\eeq
Then, to get the action \eq{S_CS}, we apply the rules \eq{rules} to 
the above expression of $\delta\Gamma$ and plug that into 
$\Omega_{\rm CS}^{(2)}$, which is not so bad because we can use the
gauge condition $h_{3A}=0$ from the beginning of the calculation.

What is grueling is to get the $c$-dependent terms in the constraint equations
\eq{const_eq_1}-\eq{const_eq_3}, because we need to keep $h_{3M}$ to linear 
order until the end of the calculation in order for us to be able to vary 
$S_{\rm CS}$ with respect to $h_{3M}$.  Fortunately, in the above expression 
\eq{Omega-2} of $\Omega_{\rm CS}^{(2)}$, we have no more than one $\nabla$
acting on $h_{AB}$, so we can still use the rules \eq{rules}.  This is a 
lengthy but straightforward calculation.  A better way is to first combine the
two terms in \eq{Omega-2} to get
\beq
  \Omega_{\rm CS}^{(2)}
  \equiv \epsilon^{ABC} \,
         {\rm Tr}\bigl[\delta{\bf \Gamma}_{\!A} \nabla_{\!B} 
                       \delta{\bf \Gamma}_C \bigr]  \>.
\eql{Omega-2-2}
\eeq
This simple appearance is actually deceiving, because now we have two 
$\nabla$'s acting on $h_{AB}$, so we need extend the rules \eq{rules} to 
the case with two covariant derivatives, which will be many more rules than 
the one-derivative case.  So, we should use the commutation relation 
\eq{nabla-commu} to eliminate $\nabla$'s as much as possible.  Below we
sketch how the calculation proceeds when one does it this way.

First, using \eq{deltaGamma}, we can write \eq{Omega-2-2} explicitly in terms
of $h_{AB}$:
\beq
  \Omega_{\rm CS}^{(2)}
  = \frac14 \epsilon^{ABC} 
      (\nabla_{\!\!A} h^{DE}) \, \nabla_{\!B} \nabla_{\!C} h_{DE}
    + \frac12 \epsilon^{ABC}
      (\nabla^E h^D_A) \, \nabla_{\!B}( \nabla_{\!D} h_{CE}
                                        - \nabla_{\!E} h_{CD} )  \>.
\eeq
Then, varying $\Omega_{\rm CS}^{(2)}$ with respect to $h_{AB}$ gives
\beq
  &&  \frac{\delta}{\delta h_{DE}} \int\! d^3x \, \Omega_{\rm CS}^{(2)}  \nn\\
  &=& -\frac12 
       \left[ 
         \frac12\epsilon^{ABC} \nabla_{\!\!A} \nabla_{\!B} \nabla_{\!C} h_{DE}
         +\epsilon^{EBC} \nabla_{\!\!A} \nabla_{\!B} 
          (\nabla^D h^A_C - \nabla^A h^D_C)
         + (D \leftrightarrow E)
       \right]
\eql{Omega-2-variation}
\eeq
where $(D \leftrightarrow E)$ represents the whole expression before it with
$D$ and $E$ swapped.  Now, the three $\nabla$'s in the first term above
can be immediately reduced to one $\nabla$ using the commutator \eq{nabla-commu}
since they are already anti-symmetrized due to the $\epsilon$ tensor.  
In a maximally symmetric space such as AdS$_3$, it simplifies down to
\beq
  \epsilon^{ABC} \nabla_{\!\!A} \nabla_{\!B} \nabla_{\!C} h_{DE}
  = -\frac{\hat{R}}{6}
       \left[ \epsilon^{DAB} \nabla_{\!\!A} h^E_B 
             +(D \leftrightarrow E) \right]  \>.
\eql{ABC}
\eeq
As for $\epsilon^{EBC} \nabla_{\!\!A} \nabla_{\!B} \nabla^D h^A_C$, 
we can simplify it (in a maximally symmetric space) as
\beq
   \epsilon^{EBC} \nabla_{\!\!A} \nabla_{\!B} \nabla^D h^A_C
   +(D \leftrightarrow E)
  = \epsilon^{EBC}
      \left( \nabla^D \nabla_{\!B} \nabla_{\!\!A} h^A_C
            + \frac{2\hat{R}}{3} \, \nabla_{\!B} h^D_C
      \right)
    +(D \leftrightarrow E)  \>.
\eeq
This is in fact better than the original expression; first, the $\hat{R}$ term
can be combined with \eq{ABC}.  Second, note that the 
$\nabla_{\!B}$ in the first term can be replaced with $\del_B$.  Then, 
we write out the $\nabla^D$ explicitly in terms of $\del$ and $\hat{\Gamma}$.
Now we have only one $\nabla$ left, which in our $h_{3A}=0$ gauge gives
\beq
  \nabla_{\!\!A} h^A_C
  = \del_A h^A_C + \frac{1}{z} \delta^3_C h  \>.
\eeq
We then obtain
\beq
 && \epsilon^{EBC} \nabla^D \nabla_{\!B} \nabla_{\!\!A} h^A_C
    +(D \leftrightarrow E)  \nn\\
 &=& \epsilon^{DAB} \hat{g}^{EF}
       \biggl[ \del_F \del_A \del_C h^C_B 
              +\frac{1}{z} \delta^3_A \del_F \del_C h^C_B
              +\frac{1}{z} \delta^3_B \del_F \del_A h
              +\frac{1}{z} \delta^3_B \del_A \del_C h^C_F  \nn\\
                 && \hspace{5EM}     
              -2z \,\hat{g}_{3F} \, \del_A \del_C h^C_B
              -2 \hat{g}_{3F} \, \delta^3_B \del_A h
       \biggr] 
     + (D \leftrightarrow E)  \>.
\eeq
Next, going back to \eq{Omega-2-variation}, we have   
\beq
   \epsilon^{EBC} \nabla_{\!\!A} \nabla_{\!B} \nabla^A h^D_C
   +(D \leftrightarrow E)
  = \epsilon^{EBC}
      \left( \nabla_{\!B} \nabla^2 h^D_C + \frac{\hat{R}}{6} \nabla_{\!B} h^D_C
      \right)
   +(D \leftrightarrow E)  \>.
\eeq
Again, the $\hat{R}$ term can be combined with \eq{ABC}.  The three-$\nabla$ term is not so bad since two of them are contracted, and exploiting the
anti-symmetry between $B$ and $C$, we can simplify it to  
\beq
   \epsilon^{EBC} \nabla_{\!B} \nabla^2 h^D_C + (D \leftrightarrow E)
  = \epsilon^{EBC}
      \left( \del_B - \frac{1}{z} \delta^3_B \right) \! \nabla^2 h^D_C  \>.
\eeq
This $\nabla^2$ term must be computed by brute force, but this is the only one.
It becomes
\beq
  \nabla^2 h^D_C
  = \hat{g}^{AF} \del_A \del_F h^D_C + z \del_3 h^D_C
   +\frac{2}{z} \delta^3_C \del_A h^{AD} - 2z \delta^D_3 \del_A h^A_C
   +2h^D_C - 2\delta^D_3 \delta^3_C h  \>. 
\eeq
Putting all the pieces together (with $\hat{R}=6$ for AdS$_3$), we obtain
\beq
  &&  -\frac{\delta}{\delta h_{DE}} \int\! d^3x \, \Omega_{\rm CS}^{(2)}  \nn\\
  &=& \frac12\epsilon^{DAB}
        \biggl[ 2\nabla_{\!\!A} h^E_B
               +\hat{g}^{EF} 
                  \biggl( \del_F \del_A \del_C h^C_B 
                         +\frac{1}{z} \delta^3_B \del_F \del_A h
                         +\frac{1}{z} \delta^3_A \del_F \del_C h^C_B
                  \biggr)  \nn\\
                    && \hspace{3EM}
               -\hat{g}^{CF}
                  \biggl( \del_C \del_F \del_A h^E_B
                         +\frac{1}{z} \delta^3_A \del_C \del_F h^E_B
                  \biggr)
               -z\del_A \del_3 h^E_B 
               -\frac{1}{z} \delta^3_B \del_A \del_C h^{CE}  \nn\\
                  && \hspace{3EM}
               -2\del_A h^E_B + \frac{2}{z} \delta^3_A h^E_B             
      \biggr] + (D \leftrightarrow E)  \>.
\eeq
%



\begin{thebibliography}{99}

\bibitem{AdS-CFT}
J.~M.~Maldacena,
  ``The large N limit of superconformal field theories and supergravity,''
  Adv.\ Theor.\ Math.\ Phys.\  {\bf 2}, 231 (1998)
  [Int.\ J.\ Theor.\ Phys.\  {\bf 38}, 1113 (1999)]
  [arXiv:hep-th/9711200];
S.~S.~Gubser, I.~R.~Klebanov and A.~M.~Polyakov,
  ``Gauge theory correlators from non-critical string theory,''
  Phys.\ Lett.\  B {\bf 428}, 105 (1998)
  [arXiv:hep-th/9802109];
E.~Witten,
  ``Anti-de Sitter space and holography,''
  Adv.\ Theor.\ Math.\ Phys.\  {\bf 2}, 253 (1998)
  [arXiv:hep-th/9802150].
  
\bibitem{BMN}
  D.~E.~Berenstein, J.~M.~Maldacena and H.~S.~Nastase,
  ``Strings in flat space and pp waves from N = 4 super Yang Mills,''
  JHEP {\bf 0204}, 013 (2002)
  [arXiv:hep-th/0202021].

\bibitem{original-AdS-QCD}
  J.~Erlich, E.~Katz, D.~T.~Son and M.~A.~Stephanov,
  ``QCD and a holographic model of hadrons,''
  Phys.\ Rev.\ Lett.\  {\bf 95}, 261602 (2005)
  [arXiv:hep-ph/0501128].

\bibitem{tensor-mesons}
  E.~Katz, A.~Lewandowski and M.~D.~Schwartz,
  ``Tensor mesons in AdS/QCD,''
  Phys.\ Rev.\  D {\bf 74}, 086004 (2006)
  [arXiv:hep-ph/0510388].

\bibitem{eta-prime}
  E.~Katz and M.~D.~Schwartz,
  ``An eta primer: Solving the U(1) problem with AdS/QCD,''
  arXiv:0705.0534 [hep-ph].

\bibitem{linear-confinement}
  A.~Karch, E.~Katz, D.~T.~Son and M.~A.~Stephanov,
  ``Linear confinement and AdS/QCD,''
  Phys.\ Rev.\  D {\bf 74}, 015005 (2006)
  [arXiv:hep-ph/0602229].

\bibitem{tHooft}
  G.~'t Hooft,
  ``A Two-Dimensional Model For Mesons,''
  Nucl.\ Phys.\  B {\bf 75}, 461 (1974).

\bibitem{Gross}
  C.~G.~Callan, N.~Coote and D.~J.~Gross,
  ``Two-Dimensional Yang-Mills Theory: A Model Of Quark Confinement,''
  Phys.\ Rev.\  D {\bf 13}, 1649 (1976).

\bibitem{Rich}
  R.~C.~Brower, W.~L.~Spence and J.~H.~Weis,
  ``Bound States And Asymptotic Limits For QCD In Two-Dimensions,''
  Phys.\ Rev.\  D {\bf 19}, 3024 (1979).

\bibitem{brodsky}
  S.~J.~Brodsky and G.~F.~de Teramond,
  arXiv:0709.2072 [hep-ph].

\bibitem{Neuberger}
  R.~Narayanan and H.~Neuberger,
  ``The quark mass dependence of the pion mass at infinite N,''
  Phys.\ Lett.\  B {\bf 616}, 76 (2005)
  [arXiv:hep-lat/0503033].
  
\bibitem{CMW-theorem}
  S.~R.~Coleman,
  ``There are no Goldstone bosons in two-dimensions,''
  Commun.\ Math.\ Phys.\  {\bf 31}, 259 (1973);
  N.~D.~Mermin and H.~Wagner,
  ``Absence of ferromagnetism or antiferromagnetism in one-dimensional or
  two-dimensional isotropic Heisenberg models,''
  Phys.\ Rev.\ Lett.\  {\bf 17}, 1133 (1966).
 
\bibitem{witten}
  E.~Witten,
  ``Chiral Symmetry, The 1/N Expansion, And The SU(N) Thirring Model,''
  Nucl.\ Phys.\  B {\bf 145}, 110 (1978).

\bibitem{Deser}
  S.~Deser, R.~Jackiw and S.~Templeton,
  ``Three-Dimensional Massive Gauge Theories,''
  Phys.\ Rev.\ Lett.\  {\bf 48} (1982) 975.

\bibitem{So-Young}
  G.~Guralnik, A.~Iorio, R.~Jackiw and S.~Y.~Pi,
  ``Dimensionally reduced gravitational Chern-Simons term and its kink,''
  Annals Phys.\  {\bf 308}, 222 (2003)
  [arXiv:hep-th/0305117].

\bibitem{DISads}
  J.~Polchinski and M.~J.~Strassler,
  ``Deep inelastic scattering and gauge/string duality,''
  JHEP {\bf 0305}, 012 (2003)
  [arXiv:hep-th/0209211].
\end{thebibliography}
\end{document}